\documentclass{JHEP3}

\usepackage{amssymb}
\usepackage{amsfonts}
\usepackage{amsbsy}
\usepackage{amsmath}
\usepackage{amsthm}
\usepackage{graphicx}
\usepackage{epstopdf}
\usepackage[vcentermath]{youngtab}
\usepackage{multirow}
\usepackage{latexsym}
\usepackage{array}

\renewcommand{\mod}[1]{\ (\mathrm{mod}\ #1)}

\def\half{\frac{1}{2}}

\def\P{\mathbb{P}}

\def\nhv{H-V}
\def\n3a{t}

\def\tr{{\mathrm{tr}}}

\newcommand{\be}{\begin{equation}}
\newcommand{\ee}{\end{equation}}
\newcommand{\eq}[1]{(\ref{#1})}
\def\ov{\over}

\title{6D supergravity without tensor multiplets}

\author{Vijay Kumar$^{1}$, Daniel S.\ Park$^2$ and Washington Taylor$^2$\\
$^1$Kavli Institute for Theoretical 
Physics\\ University of California, Santa Barbara\\ Santa Barbara, CA
93106, USA\\
$^2$Center for Theoretical Physics\\
Department of Physics\\
Massachusetts Institute of Technology\\
Cambridge, MA 02139, USA\\
\\
\\
{\tt vijayk} {\rm at} {\tt kitp.ucsb.edu},
{\tt whizpark} {\rm at} {\tt mit.edu},
{\tt wati} {\rm at} {\tt mit.edu}
}

\preprint{MIT-CTP-4190, NSF-KITP-10-139}

\abstract{We systematically investigate the finite set of possible
  gauge groups and matter content for ${\cal N} = 1$ supergravity
  theories in six dimensions with no tensor multiplets, focusing on
  nonabelian gauge groups which are a product of SU($N$) factors.  We
  identify a number of models which obey all known low-energy
  consistency conditions, but which have no known string theory
 realization.  Many of these models contain novel matter
  representations, suggesting possible new string theory
  constructions.    Many of the most exotic matter
  structures arise in models which precisely saturate the
  gravitational anomaly bound on the number of hypermultiplets.  Such
  models have a rigid symmetry structure, in the sense that there are no moduli
  which leave the full gauge group unbroken.  }

\begin{document}

\section{Introduction}

Six-dimensional theories of gravity with minimal supersymmetry provide
a rich domain in which to study fundamental questions about the space
of supersymmetric string theory vacua and consistency constraints on
low-energy supergravity theories.  Such theories can contain gauge
groups and matter in various representations, providing structure
analogous to the symmetries and particles seen in our observed
four-dimensional region of the universe.  A wide variety of string
constructions give rise to low-energy six-dimensional supergravity
theories with different gauge groups and matter content \cite{variety}.  At
the same time, we have sufficient analytic control over supergravity
in six dimensions to begin to systematically address global questions
about the space of possible theories.

In a recent series of papers \cite{universality, finite, KMT, tensors}
(summarized in \cite{singer}), it was shown that when the number of
tensor multiplets is less than 9, there are a finite number of
possible distinct gauge groups and matter content compatible with
known low-energy consistency conditions for 6D ${\cal N} = 1$
supergravity.  It was furthermore shown that the anomaly cancellation
conditions for such theories give rise to an integral lattice which
provides a direct connection to topological F-theory data for
constructing such theories.  In some cases, the resulting topological
data cannot be consistent with F-theory geometry in any known fashion,
so that a class of theories can be identified which satisfy all known
low-energy consistency conditions and yet are not realized in
conventional F-theory or any other known string construction.

In this paper we initiate a systematic analysis of the class of
theories with $T = 0$ tensor multiplets.  In this case the anomaly
cancellation conditions are particularly simple, so that a complete
solution and classification of possible low-energy models is
tractable.  We focus here on models with gauge group of the form $G =
SU(N_1) \times \cdots \times SU(N_k)$, and consider all possible
structures of matter representations.  We find that as the rank of the
gauge group factors decreases, exotic matter representations arise in
the low-energy theory.  Some of these exotic matter types are not
currently reproducible from F-theory or any other approach to string
compactification.    We give a systematic
classification of the new kinds of matter representations which can
arise in these apparently consistent low-energy theories.  The new
representations may give hints for new codimension 2 singularity
structures in F-theory.  Or they may be associated with novel
pathologies obstructing a UV completion, which may be identifiable
from the low-energy theory.

A number of specific 6D theories without tensor multiplets have
previously been constructed using particular string compactifications.
While perturbative heterotic compactifications on a K3 surface 
have a single tensor
multiplet arising from the reduction of the anti-self-dual part
of the 10D $B$ field, by passing through a tensionless string
transition the tensor multiplet can be removed from the spectrum
\cite{Seiberg-Witten, Witten, Morrison-Vafa-II}.  In F-theory this
leads to the geometrically simplest class of 6D vacuum construction,
based on an elliptic fibration over $\P^2$.  Other $T = 0$ models have
been identified using type I constructions and Gepner models
\cite{Bianchi, Schellekens}. 
The gauge groups and matter representations in these models can also
be realized through F-theory constructions of the type considered in
this paper. 
While we focus on gauge groups built from $SU(N)$ factors,
the results can be easily extended to all semi-simple Lie groups.

In Section \ref{sec:review}, we review the basic structure of
low-energy 6D supergravities, specializing to the case without tensor
multiplets.  We also review the F-theory construction of such models.
In Section \ref{sec:results}, we summarize the results of our analysis
and give explicit examples of apparently-consistent
models containing   exotic
matter representations.  Some comments and conclusions are given in Section
\ref{sec:conclusions}.  Appendices contain proofs of several technical
results used in the main text.

Note that although in this paper we refer to certain features of
low-energy theories in relation to possible F-theory constructions,
the classification of apparently consistent low-energy models is
independent of any assumptions about the UV completion of the models.
More explicit analysis of F-theory realizations of some of the models
presented here will appear elsewhere \cite{Morrison-Taylor}.

\section{6D supergravity without tensor multiplets}
\label{sec:review}

In this section we briefly review the structure of 6D supergravity
theories, specializing to the case without tensor multiplets.
Theories with an arbitrary number of tensor multiplets were developed
in \cite{Romans}, and anomaly cancellation in such models was analyzed
in \cite{Sagnotti, Sadov, tensors}, generalizing the Green-Schwarz
anomaly cancellation mechanism for theories with one tensor multiplet
described in \cite{Green-Schwarz-West, Gates-Nishino}.  We follow here
the notation and conventions of \cite{tensors}, to which the reader is
referred for further background.

\subsection{6D supergravity and anomaly cancellation}
\label{sec:anomalies}

Classical ${\cal N} = 1$ supergravity in six dimensions contains
fields in four distinct representations of the supersymmetry algebra.
Each  theory contains a single gravity multiplet, whose bosonic
components describe the metric tensor $g_{\mu \nu}$ and a self-dual
2-form field $B^+_{\mu \nu}$.  The theory can contain any number $T$
of tensor multiplets, each of which contains an anti-self-dual 2-form
field.  In this paper we specialize to the case $T = 0$.  The gravity
theory can also be coupled classically to an arbitrary gauge group $G$
described by $V$ vector fields ($V =$ dim $G$), and $H$ hypermultiplets
containing scalars transforming in an arbitrary representation of $G$.
In this paper we restrict attention to theories having gauge groups with
nonabelian structure
\begin{equation}
G = G_1 \times \cdots \times G_k
 = SU(N_1) \times \cdots
\times SU(N_k)\,.
\label{eq:group-form}
\end{equation}
A similar analysis can be done for other nonabelian gauge group
structures.  Abelian gauge group factors do not significantly modify
the story for the nonabelian part of the
theory, and will be discussed elsewhere. 

Quantum (semi-classical) consistency of supergravity theories in six dimensions
requires that anomalies cancel through the Green-Schwarz mechanism
\cite{Green-Schwarz-West, Sagnotti, Sadov}.  As described in
\cite{tensors}, in the case $T = 0$ the anomaly cancellation
conditions can be written in terms of a set of integers $b_i$
associated with the simple factors $G_i$ of the gauge group
\begin{align}
H_{\rm total}-V & =  H_{\rm neutral} + H-V  =   273 \label{eq:gravity-anomaly} \\ 
3 b_i & = \frac{1}{6} \left[    \sum_R
x^i_R A^i_R   - A^i_{\rm adj} \right]\label{eq:ab-condition}\\
0 & =   \sum_R x_R^i B^i_R  -
B^i_{\rm adj}  \label{eq:b-condition}\\
b_i^2 & =  \frac{1}{3}  \left[ \sum_R x_R^i C^i_R  -
C^i_{\rm adj} \right]  \label{eq:bb-condition}\\
b_i b_j & =  \sum_{RS} x_{RS}^{ij} A_R^i
A_S^j.
\label{eq:bij-condition}
\end{align}
In these anomaly cancellation conditions the quantities $x_R^i$ denote
the number of matter fields which transform in the irreducible
representation $R$ of gauge group factor $G_i$.  Similarly,
$x_{RS}^{ij}$ denotes the number of matter fields transforming under
representation $R \times S$ of $G_i \times G_j$.  The constants $A_R,
B_R, C_R$ are group theory coefficients defined through
\begin{align}
\tr_R F^2 & = A_R  \tr F^2 \\
\tr_R F^4 & = B_R \tr F^4+C_R (\tr F^2)^2 \,.
\end{align}
We denote by $H$ the number of matter hypermultiplets carrying
nonabelian charges and $H_{\rm neutral}$ the number of neutral
hypermultiplets.  Note that because we have specialized to models with
simple gauge group factors $SU(N_i)$, the normalization factors
$\lambda_i$ appearing in the anomaly cancellation conditions as
presented in \cite{tensors} are all unity ($\lambda_i = 1$) and do not
appear in our equations.  Note that conjugate representations $R$ and
$\bar{R}$ contribute in the same way to the anomaly conditions.  We
will not distinguish between representations and their conjugates in
our analysis here; the information about whether each matter field is
in a particular representation or its conjugate (or a linear
combination) represents an additional discrete degree of freedom which
parameterizes the full set of possible models.

In addition to local anomalies, quantum consistency requires the
absence of global anomalies \cite{Witten-SU2}.   For $T = 0$ models
with $SU(N)$ gauge group factors,
the absence of global anomalies is guaranteed for any model without
local anomalies.  This result is proven in Appendix A.

Another low-energy condition, which was used in \cite{finite, tensors}
to prove that the number of gauge groups and matter representations
associated with consistent theories is finite for $T < 9$, is the
constraint that all gauge kinetic terms have the proper sign
\cite{Sagnotti}.  In theories with $T = 0$ this is simply the
constraint that all $b_i$ have the same sign.  As we demonstrate below,
with the sign conventions chosen here, there are no models consistent
with anomaly cancellation which have $b_i \leq 0$, so the gauge
kinetic term sign condition is automatically satisfied.

In a general 6D supergravity theory, the tensor multiplet moduli define the coupling constants, or the strength of the gauge interactions relative to gravity. Theories with $T=0$ are, therefore, intrinsically gravitational with all interaction strengths set by the Planck scale. 

\subsection{F-theory realizations of $T = 0$  6D models}

F-theory \cite{Vafa-f, Morrison-Vafa-I, Morrison-Vafa-II} is a very
general approach to constructing string vacua in even dimensions.
F-theory is particularly useful
in describing models without tensor multiplets.  We briefly summarize
the basic aspects of F-theory realizations of 6D theories here,
specializing to the case $T = 0$.  The F-theory picture will not be
used in deriving the results in the remainder of the paper, so readers
without background in this area can skip this section if they like.
The main result we take from this summary is the condition
(\ref{eq:Kodaira}) which places a bound on the range of possible
low-energy models with F-theory realizations.  A more detailed
discussion of F-theory constructions and the correspondence with
low-energy 6D supergravity models is given in \cite{tensors}.

F-theory models in six dimensions are constructed using a Calabi-Yau
threefold which admits an elliptic fibration with section.  For $T =
0$, the base $B$ of the elliptic fibration is just complex projective
space $\P^2$.  This is thus the simplest class of F-theory
compactifications below 8 dimensions.  The gauge group of an F-theory
compactification is determined by the codimension one singularities in
the fibration.  The Kodaira type of each singularity determines an
associated nonabelian gauge group factor in a way which is now well
understood \cite{Morrison-Vafa-I, Morrison-Vafa-II}.  When
compactifying to 6 space-time dimensions, the elliptic
fibration can develop further codimension two singularities.  These
correspond to matter transforming under various representations of the
gauge group.  Some types of codimension two singularities have been
analyzed and associated to specific types of matter representations
\cite{Bershadsky-all, Sadov, Katz-Vafa, Grassi-Morrison, Grassi-Morrison-2}.  A
complete analysis of codimension two singularities is still lacking,
however, and  the complete range of possible matter
representations which can be realized in F-theory is not yet systematically
understood.

The discriminant locus of an elliptic fibration is given by a
divisor class $\Delta$ in the base $B$.  This can be decomposed into a
sum of components
\begin{equation}
\Delta = \sum_{i}N_i \xi_i + Y
\end{equation}
where $\xi_i$ are irreducible effective divisors giving rise to
nonabelian gauge factors and $Y$ is a residual effective divisor.  For
the gauge groups we are considering here, with all simple factors of
the form $SU(N_i)$, for example, $\xi_i$ corresponds to an $A_{N_i
  -1}$ singularity for $N_i > 3$ (there are several possible ways of
realizing the groups $SU(2)$ and $SU(3)$).  The structure of the group of divisors on $\P^2$ is very simple.  All divisors
are integer multiples of the hyperplane divisor $H$, so $\xi_i =
\beta_iH$ with $\beta_i > 0$ for irreducible effective divisors.  The
canonical class of $\P^2$ is $K = -3H$.  The Kodaira
condition stating that the total space of the elliptic fibration is a
Calabi-Yau manifold is
\begin{equation}
-12K = 36H = \Delta =\sum_{i}N_i \xi_i + Y \,.
\end{equation}

As shown in \cite{KMT, tensors}, the correspondence between components
of the discriminant locus
and the  anomaly structure \cite{Sadov, Grassi-Morrison}
can be used to construct a map from any given low-energy theory to
topological data for an F-theory fibration.  
For $\P^2$, this map is uniquely defined and quite simple.
Associated with the coefficient $b_i$ for each simple factor in the
gauge group there is a divisor $\xi_i = b_iH$, so the map is
\begin{equation}
b_i \rightarrow \beta_i \,.
\label{eq:map}
\end{equation}
{}From this we see that low-energy theories arising from F-theory must
satisfy several constraints.  Because $\xi_i$ must be effective, we have
\begin{equation}
b_i > 0 \,.
\label{eq:positivity}
\end{equation}
In \cite{tensors} a more general version of this condition is
described, corresponding to the constraint that the divisors $\xi_i$
must lie in the Mori cone.  [Note that in writing the anomaly
  condition (\ref{eq:ab-condition}), from the point of view of the
  more general formalism of \cite{tensors}, we have chosen a sign for
  the gravitational anomaly coefficient $a$ (corresponding to the
  low-energy manifestation of the canonical class $K = -3H$) which
  leads to the sign of the constraint (\ref{eq:positivity})].

The Kodaira constraint gives the upper bound
\begin{equation}
36 \geq \sum_{i} N_i b_i \,,
 \label{eq:Kodaira}
\end{equation}
since the residual divisor locus $Y$ must also be effective.  As we
will see, the positivity condition (\ref{eq:positivity}) is
automatically satisfied for any low-energy theory, but the Kodaira
condition (\ref{eq:Kodaira}) provides a criterion for showing that
some apparently consistent low-energy theories cannot be realized in
F-theory as it is currently understood.  In \cite{tensors} some other
F-theory constraints on low-energy theories are described, such as the
condition that the anomaly coefficients live in a unimodular lattice;
these other constraints are automatically satisfied for models with $T
= 0$.

\section{Possible models and exotic matter representations}
\label{sec:results}

In this section we begin with some general observations regarding the
structure of $T = 0$ 6D models and then present the results of a
systematic analysis of those models with gauge group of the form
(\ref{eq:group-form}).

\subsection{Block decomposition of models}

In systematically determining what kinds of models are possible, we
can use the fact that the anomaly equations depend primarily on the
integers $b_i$ associated with each gauge group factor separately.
Only the cross-term component (\ref{eq:bij-condition}) of the anomaly
factorization condition depends upon more than one distinct $b_i$.  By
using the other anomaly conditions we can constrain the gauge group
factors $SU(N_i)$ and matter transforming under each factor
independently.  We can then treat these factors and associated matter
as ``blocks'' which can be combined to build models with multiple
gauge group factors.  This general approach is discussed in \cite{KMT}
and used there to construct $T = 1$ models with gauge groups which are
products of $SU(N)$ factors with a restricted class of
representations.

For models with $T = 0$ the classification of blocks is particularly
simple; the integer $b_i$ associated with each factor $SU(N_i)$ in the
gauge group places strong constraints on allowed representations.  To
see how the possible representation content is structured it is
helpful to go into slightly more detail regarding the properties of
the group theory coefficients $A_R, B_R, C_R$.  As discussed for
example in \cite{finite} (see also \cite{Erler}), these group theory
coefficients can be computed for any particular representation using
two diagonal generators $T_{12}, \ T_{34}$ which, in the fundamental
representation, take the form
\begin{align}
(T_{12})_{ab} &= \delta_{a1}\delta_{b1}-\delta_{a2}\delta_{b2} \\
(T_{34})_{ab} & = \delta_{a3}\delta_{b3} - \delta_{a4}\delta_{b4}
\end{align}
The group theory factors $A_R, \ B_R, \ C_R$ can be computed in terms
of traces of these generators.
For $SU(N), N > 3$, we have
\begin{align}
A_R & = \frac{1}{2} \tr_R T_{12}^2  \label{eq:A-def}\\
B_R+2 C_R & = \half \tr_R T_{12}^4 \label{eq:B-def}\\
C_R & = \frac{3}{4} \tr_R T_{12}^2T_{34}^2 \label{eq:C-def}
\end{align}
In these traces, we sum over all basis states in the 
representation $R$, which can be represented in terms of the Young
tableaux with various labelings of the
associated Young diagram $D_R$.  
For $SU(2)$ and $SU(3)$ there is no fourth order Casimir, or generator
$T_{34}$, so we can take $B_R = 0$ and use (\ref{eq:B-def}) to compute
$C_R$.  We will find it useful to work with the linear combination
\begin{equation}
g_R := \frac{1}{12}\left(2 C_R + B_R -A_R \right) = \frac{1}{24} \left( \tr_R T_{12}^4 - \tr_R T_{12}^2 \right)\,.
\end{equation}
Since in any given state in the representation $T_{12}^2 \leq
T_{12}^4$, we see that 
\begin{equation}
g_R \geq 0, \;\;\;\;\; \forall R \,.
\end{equation}
For representations given
by Young diagrams with a single column there are no states with $| \langle
T_{12} \rangle | > 1$ and therefore $g_R = 0$; all other representations have
$g_R > 0$.  

For a gauge group factor $SU(N)$ with corresponding
anomaly integer $b$, we can take a linear combination of the anomaly conditions (\ref{eq:ab-condition}), (\ref{eq:b-condition}), (\ref{eq:bb-condition}) to get
\begin{equation}
\sum_{R}x_R \, g_R =  \frac{1}{2} \left(2g_{\rm adj} +b^2 -3b \right) = \frac{(b-1) (b-2)}{2}
\label{eq:genus-equation}
\end{equation}
where we have used $g_{\rm adj} = C_{\rm adj}/6 = 1$.  For models with
an F-theory construction, the anomaly integer $b$ is the degree of the
curve realizing the corresponding gauge group.  
The quantity $g :=
\sum_R x_R g_R = (b-1) (b-2)/2$ is then the (arithmetic)
genus of this curve.  We thus refer to $g_R$
as the ``genus'' of representation $R$, anticipating that for
situations with an F-theory realization it will have this geometric
interpretation.  
In F-theory,
the number of adjoint hypermultiplets in the
low-energy theory is given by the geometric genus $g_g$ of the curve.
The genus-degree formula for a general, possibly singular,  curve relates the arithmetic
and geometric genera
\begin{equation}
g =  (b-1) (b-2)/2 = g_g + \sum_P \frac{m_P(m_P-1)}{2},
\end{equation}
where the sum is over all singular points $P$ of the curve, and $m_P$ is the
multiplicity at point $P$ \cite{perrin}.  This relationship provides a
clue towards realizing general matter representations in F-theory
through codimension-2 singularities.
This point will be explored
further in \cite{Morrison-Taylor}.

Some examples of group theory coefficients, dimensions, and genera are
shown in Table~\ref{t:coefficients}.



\begin{table}
\centering
\begin{tabular}{|c|c|c|c|c|c|c|}
\hline
 Rep. & Dimension & $A_R$ & $B_R$ & $C_R$  & $g_R$\\
\hline
 ${\tiny\yng(1)}$ & $N$ & 1 & 1 & 0 &  0\\
 Adjoint & $N^2-1$ & $2N$ & $2N$ & 6 & 1\\
 ${\tiny\yng(1,1)}$ & $ \frac{N(N-1)}{2} $ & $ N-2 $ & $ N-8 $ & 3 &  0\\
 ${\tiny\yng(2)}$ & $ \frac{N(N+1)}{2} $ & $ N+2 $ & $ N+8 $ & 3 & 1\\
 ${\tiny\yng(1,1,1)}$ & $ \frac{N(N-1)(N-2)}{6} $& $
 \frac{N^2-5N+6}{2} $&$ \frac{N^2-17N+54}{2} $ & $ 3N-12 $ &  0\\
 ${\tiny\yng(2,1)}$ & $ \frac{N(N^2-1)}{3} $& $ N^2-3 $& $ N^2-27 $& $
 6N $ &  $N -2 $\\
 ${\tiny\yng(3)}$ & $ \frac{N(N+1)(N+2)}{6} $ &$ \frac{N^2+5N+6}{2}
 $&$ \frac{N^2+17N+54}{2} $ & $ 3N+12 $ &  $N + 4$\\
 ${\tiny\yng(1,1,1,1)}$ & $ \frac{N(N-1)(N-2)(N-3)}{24} $&
 $ \frac{(N-2)(N-3)(N-4)}{6} $ & $\frac{(N-4)(N^2-23N+96)}{6}$ & 
$\frac{3 (N^2 -9 N + 20)}{2}$  &  0\\
 ${\tiny\yng(2,2)}$ &  $ \frac{N^2(N+1)(N-1)}{12} $ &
 $\frac{N(N-2)(N+2)}{3} $ & $\frac{N(N^2-58)}{3}$ & $3(N^2+2)$  &  $\frac{(N-1)(N-2)}{2}$ \\[0.07in]
\hline
\end{tabular}
\caption{Values of the group-theoretic coefficients $A_R, B_R, C_R$,
  dimension and genus
for some representations of $SU(N)$, $N \geq 4$.
For $SU(2)$ and $SU(3)$, $A_R$ is given in table, while
$B_R = 0$ and $C_R$ is computed by adding formulae for $C_R + B_R/2$
from table with $N = 2, 3$.}
\label{t:coefficients}
\end{table}

We can now easily show that there are no $SU(N)$ factors in the gauge
group with $b < 0$.  For such a factor, from 
(\ref{eq:genus-equation})
we have
\begin{equation}
\sum_{R} x_R g_R=\frac{1}{2}\left(2 + b^2 -3b\right) \geq 3 \,,
\end{equation}
so some representations with nonzero genus must be included.  From
(\ref{eq:ab-condition})
we have
\begin{equation}
A_{\rm adj} +18b = 2 N - 18 | b | = \sum_{R}x_R A_R \,.
\end{equation}
Since all $A_R$ are positive, this implies $N > 9$.  But for $N > 9$ a matter hypermultiplet in any 
representation $R$ with $g_R > 0$ satisfies $x_R A_R > 2 N -18$, with the exception of a single hypermultiplet in the 2-index symmetric representation ${\tiny\yng(2)}$.
In that case, however, we would have $ \sum_{R} x_R g_R = 1 < 3$, and so there are
no gauge group factors with $b < 0$.

\subsection{Single factors}

We can now address the question of what group factors $SU(N)$ and
associated representations can appear in a low-energy supergravity
theory satisfying the anomaly constraints.  Unlike the situation for
theories with $T > 0$ tensor fields, for $T = 0$ theories the fact
that there are a finite number of possible ``blocks'' associated with
gauge group factors $SU(N)$ and particular matter representations
follows without using the gravitational anomaly bound
(\ref{eq:gravity-anomaly}).  This result is proven in
Appendix B.  The consequence of the finiteness of the set of blocks,
independent of  any bound on the number of hypermultiplets in the
matter representations, is that we can simply enumerate all possible
gauge group blocks.  We can then in principle figure out all ways of
combining these blocks to form low-energy models consistent with
anomaly cancellation.

One conceptually straightforward way to see how to perform  an
enumeration of individual blocks is as follows.  Fix $N$ and $b$.  Then
(\ref{eq:genus-equation}) gives a bound on the sum of the non-negative
values $g_R$ associated with the matter representations transforming
under $SU(N)$.  This gives a finite partition problem, to which all
solutions can be found.  Each solution of the partition problem
corresponds to a set of values for the $x_R$ associated with
representations with nonzero genus $g_R$.
As noted above, the representations with $g_R>0$ are all
associated with Young diagrams having more than one
column.  We can then fix the $x_R$
for all $R$ with $g_R > 0$
and treat (\ref{eq:bb-condition}) as a second
partition problem.  Since all $C_R$ are positive except for the
fundamental representation, this gives a set of possible combinations of
coefficients $x_R$ for all representations besides the fundamental.
We can then use (\ref{eq:b-condition}) to determine the number of
fundamental representations, which must be nonnegative.
\vspace*{0.05in}

As an example of how this analysis works we begin by considering the
set of blocks with $b = 1$.  From (\ref{eq:genus-equation}) we have
\begin{equation}
b = 1: \;\;\;\;\; \;\;\;\;\;
2 \sum_{R}x_R\,  g_R  = (b-1) (b-2) = 0 \,.
\end{equation}
Thus, $x_R= 0$ for any representation with $g_R > 0$, and
we cannot include any representations other than those with a
single column.  The anomaly condition (\ref{eq:bb-condition}) becomes
\begin{equation}
 \sum_{R}x_R\, C_R = 9 \,.
\end{equation}
For $N > 7$, the coefficients $C_R$ satisfy $C_R > 9$ for all
one-column representations other than the two-index antisymmetric (A2)
and fundamental (F) representations.  So in these cases the only
solution is $x_{A2} = 3$.  The anomaly condition (\ref{eq:b-condition})
then becomes
\begin{equation}
\sum_{R}x_R\, B_R = x_F +  3(N -8) = B_{\rm adj} = 2 N \,,
\end{equation}
so
\begin{equation}
 x_F = 24 -N \,.
\end{equation}
Thus, for $b = 1$ there are no possible blocks with $N > 24$, and the
only possible blocks with $N > 7$ are $SU(N)$ factors with matter
content
\begin{equation}
(24 -N) \times {\tiny\yng(1)} +
3 \times {\tiny\yng(1,1)}, \;\;\;\;\; (b = 1, N \leq 24, H-V = (2 + 45
N -N^2)/2\leq 273) \,.
\label{eq:b1}
\end{equation}
(Recall that when describing the hypermultiplet matter content of any
block or model we denote by $H$ the number of matter
hypermultiplets which carry nonabelian charges; as long as this
quantity satisfies $H-V \leq 273$, uncharged hypermultiplets can be
added to saturate the gravitational anomaly condition
(\ref{eq:gravity-anomaly}).)

For $N \leq 7$, other  $b = 1$ blocks are possible.
It is easy to verify that including the 3-antisymmetric (A3)
representation at the second step of the above analysis for $SU(7)$
gives a block
satisfying the anomaly cancellation conditions with
\begin{equation}
SU(7): \;\;\;\;\; \;\;\;\;\;
22 \times {\tiny\yng(1)} +
1 \times {\tiny\yng(1,1,1)}, \;\;\;\;\; (b = 1, H-V = 141) \,.
\label{eq:b1b}
\end{equation}
A similar block can be constructed for $SU(6)$ with 20 fundamental, one
A2, and one A3 representation.
Since for $SU(5)$ the A3 and A2 representations are conjugate (and
therefore treated as equivalent in this analysis), this
exhausts the range of possibilities for $b = 1$.
Note that all these blocks automatically satisfy the gravitational
anomaly bound $H-V \leq 273$.
\vspace*{0.05in}

A similar analysis for $b = 2$ again allows only single-column
representations, which now restrict $N \leq 12$ and includes  $SU(N)$
blocks of the form
\begin{equation}
(48-4 N) \times {\tiny\yng(1)} +
6 \times {\tiny\yng(1,1)}, \;\;\;\;\; (b = 2, N \leq 12, H-V = 1 + 45
N -2 N^2 \leq 273) 
\label{eq:b2}
\end{equation}
for all $N \leq 12$.
Other $b = 2$ blocks are possible for  $6 \leq N \leq 10$: 
blocks with single 3-antisymmetric (A3) representations are possible at $N = 10,
9$ with $H-V > 273$ and at $N = 8, 7, 6$ with $H-V \leq 273$.
For $SU(6)$ there are also blocks with two and three A3 representations,
and for $SU(7)$ there is a block with two A3 representations; all
these blocks satisfy the gravitational anomaly bound $H-V  \leq 273$.
There is also a single $b = 2$ block with gauge group $SU(8)$ and a
4-antisymmetric (A4) representation
\begin{equation}
SU(8): \;\;\;\;\; \;\;\;\;\;
32 \times {\tiny\yng(1)} +
1 \times {\tiny\yng(1,1,1,1)}, \;\;\;\;\; (b = 2, H-V = 263) \,.
\label{eq:b2b}
\end{equation}
This exhausts the range of possibilities for $b = 2$ blocks.

Continuing to $b = 3$, there is now a nonzero contribution to the
genus,
\begin{equation}
b = 3: \;\;\;\;\; \;\;\;\;\;
2 \sum_{R}x_R \, g_R  = (b-1) (b-2) =  2 \,.
\end{equation}
There is, therefore, necessarily a matter representation with more than
one column, which has $g_R  = 1$.  The only possibilities are the
adjoint and two-index symmetric representations for general $N$
(note that the
representation ${\tiny\yng(2,1)}$ in Table~\ref{t:coefficients}
for $SU(3)$ has
$g_R = 1$, but is  also the adjoint of $SU(3)$).
For each choice of representation saturating the genus $g = 1$, there
are various possible combinations of $n$-antisymmetric single-column
representations which can solve the partition problem for the $C$'s.
The largest $N$ for which a one-block model appears with $b = 3$
which satisfies the
gravitational anomaly bound on the number of hypermultiplets is $N =
9$; the matter content of this model is
\begin{equation}
SU(9): \;\;\;\;\; \;\;\;\;\;
5 \times {\tiny\yng(1)} +
4 \times {\tiny\yng(1,1)}+
1 \times {\tiny\yng(1,1,1)}+
1 \times {\rm Adj}, \;\;\;\;\; (b = 3, H-V = 273) \,.
\label{eq:b39}
\end{equation}

Note that the blocks listed explicitly above (\ref{eq:b1},
\ref{eq:b1b}, \ref{eq:b2}, \ref{eq:b2b}, \ref{eq:b39}) have $H-V \leq 273$, and
therefore, by adding neutral hypermultiplets, can be completed to
anomaly-free low-energy supergravity theories with single factor gauge
groups $G = SU(N)$.  The model in (\ref{eq:b39}) precisely saturates
the gravitational anomaly bound with $H-V = 273$.  This model
therefore has no neutral hypermultiplets and is ``rigid'' in the sense
that deformation along any scalar modulus will break the symmetry of
the model.  As we will see, many of the most exotic matter
representations arise in such rigid models.

All the models described above furthermore satisfy the Kodaira bound
from F-theory $ \sum_{i}b_i N_i = b N \leq 36$.  We might therefore
expect that these models have F-theory realizations.  While the
fundamental and antisymmetric matter representations have standard
F-theory realizations, however, the 3-index and 4-index
representations are more exotic.  These representations were also
encountered in $T = 1$ models in \cite{KMT}.  In the case of the
3-index representations, a codimension two singularity structure has
been identified in F-theory which realizes this matter representation
for $N = 6, 7, 8$ \cite{Grassi-Morrison-2} through local enhancement
of the singularity type to $E_6, E_7$ and $E_8$ respectively.  We are
not aware, however, of any known F-theory realization of the 4-index
antisymmetric representation, or of the 3-index antisymmetric
representation for $N = 9$.

We have systematically analyzed the set of all possible $SU(N)$ blocks
with arbitrary matter representations for $T = 0$ and any $N$.  A
summary of the results of this analysis appears in
Table~\ref{t:blocks}.  We carried out this analysis by finding all of
the finite number of solutions for the partition problem for each
combination of $N$ and $b$, within the bounded range of $b$'s for
which a solution can be found for each $N$.  For $N \geq 4$ we have
explicitly computed all blocks, dividing the set into those which do
or do not individually satisfy the gravitational anomaly bound $H-V
\leq 273$.  For $SU(2)$ and $SU(3)$, the total number of blocks
becomes quite large.  For $SU(3)$ we have only explicitly computed the
number of blocks which individually satisfy the gravitational anomaly
bound, and for $SU(2)$ we have only estimated the number of blocks and
their range and computed some specific examples, as described below.
The detailed analysis of upper bounds on $b$ for each fixed $N$ is
given in the Appendix.

\begin{table}
\centering
\begin{tabular}{|c|c|c|c|c|}
\hline
N &  max $b$ & (total blocks) & \# $SU(N)$ models & \# satisfy Kodaira\\
\hline
13-24 & 1 (1) & (1) & 1 & 1\\
12 & 2  (2) &  (2) & 2 & 2\\
11 & 2  (3) &  (4) & 2 & 2\\
10 & 2  (4) &  (6) & 2 & 2\\
9 & 3  (4) &  (8) & 3 & 3\\
8 & 8  (8) &  (22) & 15 & 14\\
7 & 4  (7) &  (28) & 16 & 16\\
6 & 6  (8) &  (147) & 48 & 47\\
5 & 8  (14) &  (186) & 23 & 16\\
4 & 16  (34)  & (3893) & 207 & 154\\
3 & 597 (597) &  & 10100 & 262 \\
2 & 24297  $\leq b_{\rm max} < 36647 $ & & $\sim 5\times 10^{7}$ & 176\\
\hline
\end{tabular}
\caption{Summary of possible distinct matter representations for gauge
  group factors $SU(N)$.  Numbers in parentheses refer to possible
  blocks without constraint on number of hypermultiplets, numbers
  without parentheses refer to possible anomaly-free models with
  single nonabelian factor in total gauge group $SU(N)$.  Last column
  gives number of single factor models which satisfy Kodaira
  constraint $b N \leq 36$ needed for F-theory realization.  Number of
  blocks not individually satisfying gravitational anomaly bound
  becomes very large at $N = 3$, as does number of blocks for $N = 2$
  even with gravitational anomaly constraint.  We have not precisely
  computed the number of blocks in these categories.}
\label{t:blocks}
\end{table}  

We now describe briefly a few interesting aspects of the results
summarized in Table~\ref{t:blocks} and highlight a few specific blocks
of interest.
\vspace*{0.05in}

\noindent
{$\boldmath N > 8$}:

For $N > 9$, there are known F-theory realizations of all matter
representations appearing in all single-block models.
Furthermore, the Kodaira constraint is satisfied for all single blocks
with $N \geq 8$.  Thus, it seems likely that
all the single-block $SU(N)$ models with $N > 9$ which are
anomaly-free can be realized in F-theory.  The only unusual
representation which arises at $N = 9$ is the 3-index antisymmetric
representation mentioned above in the model (\ref{eq:b39}).
\vspace*{0.05in}

\noindent
{$\boldmath N = 8$}:

At $N = 8$ we find several novel features.  As mentioned above,
there is an $SU(8)$ model with a 4-index antisymmetric
representation.  There is also a somewhat exotic model with
\begin{equation}
SU(8): \;\;\;\;\; \;\;\;\;\;
1 \times {\tiny\yng(2,2)}  \;\;\;\;\; (b = 8, H-V = 273) \,.
\label{eq:88}
\end{equation}
This is the only $SU(8)$ model containing a block with $b > 4$ and is
another example of a model with rigid symmetry.  There is no known
F-theory realization of the ``box'' matter representation appearing in
this model.  Furthermore, this model violates the Kodaira condition
($b N = 64 > 32$).  Nonetheless, the numerology seems to work out
rather nicely for this model, suggesting that there may possibly be
some new class of string compactification which could realize this
model.
\vspace*{0.05in}

\noindent
{$\boldmath N \leq 6$}

At $N = 6$ and below, the range of possible representations expands
significantly, and
models which violate the Kodaira
condition begin to proliferate.
There is one model at $N = 6$
which has another exotic representation
\begin{equation}
SU(6): \;\;\;\;\; \;\;\;\;\;
2 \times {\tiny\yng(1,1)}+
2 \times {\tiny\yng(1,1,1)}+
2 \times {\tiny\yng(2,1,1,1)}+
2 \times {\rm Adj}, \;\;\;\;\; (b = 6, H-V = 273) \,.
\end{equation}
This is another example of a model with rigid symmetry, although this
model is (just barely) within the Kodaira bound.

At $N = 5$ and below an increasing range of exotic representations
becomes possible.  At the end of this section we summarize the set of
representations which can be realized in models satisfying the Kodaira
condition for any $N$.  
One particularly simple and interesting block with $N = 4$ is
\begin{equation}
SU(4): \;\;\;\;\; \;\;\;\;\;
1 \times {\tiny\yng(2,2)}+
64 \times {\tiny\yng(1)},  \;\;\;\;\; (b = 4, H-V = 261) \,.
\label{eq:4-box}
\end{equation}

For models not satisfying the Kodaira bound, an even wider
range of representations can be realized; for example, for $N = 4$
there are single block models violating the Kodaira bound which have
the representations ${\tiny\yng(4)}$
and ${\tiny\yng(3,2,1)}$.
Most of these exotic representations appear in models which 
precisely saturate or almost saturate the gravitational anomaly
bound.  For example, one $SU(4)$ model at $b = 16$ has
\begin{equation}
SU(4): \;\;\;\;\; \;\;\;\;\;
3 \times {\tiny\yng(3,2,1)}+
3 \times {\tiny\yng(2,2)}+
1 \times {\tiny\yng(4)},  \;\;\;\;\; (b = 16, H-V = 272) \,.
\end{equation}

At $N = 3$ the range of possibilities increases still further.  The
distribution of blocks across values of $b$ is rather non-uniform.
There are an enormous range of blocks not satisfying the gravitational
anomaly bound and having $b < 500$ which we have not attempted to
completely enumerate.  Among those blocks individually satisfying the
gravitational anomaly bound, most are distributed across values of $b
< 70$, with more blocks at values of $b$ divisible by 3.  The most
blocks satisfying the gravitational anomaly bound occur at $b = 24$
(910 blocks).  There are only a few values of $b > 70$ with allowed
such blocks, including 44 blocks at $b = 93$, followed by 3 blocks at
$b = 105$ and single blocks each at $b = 153, 168, 408$ and $597$. The
matter content for $b=597$ is given by
\begin{equation}
SU(3): \;\;\;\;\; \;\;\;\;\;
1 \times {\tiny\yng(6)} \;({\rm S6})
+1 \times \; ({\rm S21})
\;\;\;\;\; (b = 597, H-V = 273) \,.
\label{eq:597}
\end{equation}
Even without imposing the gravitational anomaly bound, there are only
blocks possible for three distinct values of $b > 500$.  At $b = 521$
there are 79,151 different blocks possible with $H-V \leq 1000$; at $b
= 522$ there are 40 such blocks.  The only block possible with $b >
522$ is (\ref{eq:597}).  It is striking that the largest possible
$SU(3)$ block precisely saturates the gravitational anomaly bound.

For $SU(2)$ we have not computed all blocks explicitly, even
restricting to blocks satisfying the gravitational anomaly bound, as
the number of possibilities is very large.  The best upper bound we
have found for $b$ for $SU(2)$ is 36,647 (see Appendix B).  We have
sampled the distribution by computing the number of blocks satisfying
the gravitational anomaly bound for multiples of 20, $b = 20k$, up to
$b = 1000$, and for multiples of 250 up to $b = 20,000$.  The number
of blocks at fixed $b$ seems to peak around $b = 420$, where there are
65,459 distinct $SU(2)$ blocks.  The number of blocks starts to drop
significantly after $b = 1000$, with for example 11,121 blocks at $b =
1000$, 835 blocks at $b = 2000$, and 12 blocks at $b = 4000$.  As for
$N = 3$, however, there are individual blocks out to much larger
values of $b$.  We have found blocks satisfying $H-V \leq 273$ for $b$
up to 24,297, though there are probably sporadic blocks appearing for
larger $b$ up to close to the bound of 36,647 (though these must be
rare; for example 24,297 is the only value of $b$ between 24,000 and
25,500 which admits a block).  Based on the partial data we have
computed, we estimate the number of blocks satisfying the
gravitational anomaly bound to be on the order of $5\times 10^{7}$.
The total number of blocks without imposing the gravitational anomaly
constraint is much larger, but still finite.  An example of an $SU(2)$
block with a very large value of $b$ satisfying the gravitational
anomaly bound is the following block with $b = 10,750$
\begin{eqnarray}
SU(2): \;\;\;\;\; \;\;\;\;\;
& & 1 \times {\tiny\yng(2)}({\rm S2})
+ 1 \times({\rm S3})+1 \times({\rm S4})
+1 \times({\rm S5})+1 \times({\rm S6}) \\
& &\hspace*{0.1in}
+1 \times({\rm S17})+1 \times({\rm S55})
+1 \times({\rm S69})+1 \times({\rm S85}), \nonumber \\
& &\hspace*{0.1in}
\;\;\;\;\; (b = 10750, H-V =252 ) \,. \nonumber 
\end{eqnarray}
An example of a block with larger $b$ which violates the gravitational
anomaly bound is
\begin{eqnarray}
SU(2): \;\;\;\;\; \;\;\;\;\;
& & 24530 \times {\tiny\yng(1)}+
8380 \times {\tiny\yng(2)}
+ 1 \times({\rm S12})+1 \times({\rm S29})
+1 \times({\rm S43})
\\& &\hspace*{0.1in}
+1 \times({\rm S113}),
\;\;\;\;\; (b = 18000, H-V =74398 ) \,. \nonumber 
\end{eqnarray}
This block, in fact, wildly violates the gravitational anomaly bound,
and it can be shown fairly easily that no model satisfying the
gravitational anomaly bound can contain this block.  For $SU(2)$ there
are many such single blocks at large $b$ that satisfy the single
block anomaly equations but violate the gravitational anomaly bound.
Thus, as the rank decreases the gravitational anomaly bound becomes a
more important constraint in restraining the class of allowed models,
even though the gravitational anomaly bound alone is sufficient to
prove that the number of blocks is finite.
\vspace*{0.1in}

We conclude this description of single $SU(N)$ factor matter blocks in
$T = 0$ models with a brief summary of all novel representations which
can appear in single block models satisfying the F-theory Kodaira
constraint, but for which no F-theory realization is known.  There is
no argument we are aware of which rules out these representations in
F-theory; indeed it seems likely that some of these representations
can be realized by new codimension 2 singular structures.  A more
detailed F-theory analysis of such singularity types will be
considered elsewhere.  Note that further representations can appear
when multiple blocks are considered, so this list is not a complete
list of all possible matter types for $T = 0$ models.

Matter representations with standard F-theory constructions are the
fundamental (${\tiny\yng(1)}$), 2-antisymmetric (A2 =
${\tiny\yng(1,1)}$), and adjoint representations
\cite{Morrison-Vafa-II}.  The 2-symmetric (S2 = ${\tiny\yng(2)}$) was
identified in terms of a double point singularity in F-theory in
\cite{Sadov} and the local singularity structure associated with
3-antisymmetric representations (${\tiny\yng(1,1,1)}$) 
have also been identified in F-theory
for $SU(6)$ \cite{Bershadsky-all, Katz-Vafa, Grassi-Morrison-2},
$SU(7)$ \cite{Katz-Vafa, Grassi-Morrison-2}, and $SU(8)$ 
\cite{Grassi-Morrison-2}.

The novel matter representations which can appear in a
model  satisfying the Kodaira
constraint,
where the gauge group has a
single nonabelian factor $SU(N)$ are as follows

\vspace*{0.05in}\noindent ${\tiny\yng(1,1,1)}$: Appears for $SU(N)$, $N =
9, 8, 7, 6$.    
\vspace*{0.05in}

\noindent ${\tiny\yng(1,1,1,1)}$: Appears for $SU(8)$
as in the single block model (\ref{eq:b2b})
\vspace*{0.05in}

\noindent
${\tiny\yng(2,1)}$: Appears for $SU(N)$, $N = 5, 4$
(Adjoint for $SU(3)$).
\vspace*{0.05in}

\noindent
${\tiny\yng(2,1,1)}$: Appears for $SU(5)$ (Adjoint for $SU(4)$).

\vspace*{0.05in}

\noindent
${\tiny\yng(2,2)}$: Appears for $SU(4)$.
\vspace*{0.05in}

\noindent
${\tiny\yng(3)}$: Appears for $SU(N)$, $N =  4, 3, 2$.

\vspace*{0.05in}

\noindent
${\tiny\yng(3,1,1)}$: Appears for $SU(4)$.

\vspace*{0.05in}

\noindent
${\tiny\yng(3,1)}$: Appears for $SU(3)$.

\vspace*{0.05in}

\noindent
${\tiny\yng(4)}$: Appears for $SU(2)$.
\vspace*{0.05in}

\noindent
${\tiny\yng(5)}$: Appears for $SU(2)$.

\subsection{Two-factor combinations}

In principle, given the complete list of all possible single blocks
one can construct all multi-block models satisfying the gravitational
anomaly bound by simply considering all possible ways in which matter
can be multiply charged between blocks in a fashion compatible with
(\ref{eq:bij-condition}).  Since the number of jointly charged
hypermultiplets grows
quickly as the number of blocks increases, the ways of combining
multiple blocks are actually quite constrained.  We have used the
complete analysis of single blocks to construct in this fashion all
possible two-block models with gauge group $SU(N) \times SU(M)$ for $4
\leq N \leq M$.  We present here some examples of the features which
can appear in such two-block models.

{}From the cross-term anomaly constraint (\ref{eq:bij-condition}), it
follows that any pair of blocks must share matter which transforms
under each gauge group factor, satisfying the summation relation
\begin{equation}
b_i b_j =  \sum_{RS} x_{RS}^{ij} A_R^i
A_S^j\,.
\end{equation}
The simplest type of matter charged under two gauge group factors is
bifundamental matter, familiar from various string constructions.  In
this case $A_R^i = A_S^j = 1$.  There is a simple family of two-block
models with matter content of the form
\begin{eqnarray}
G & = &  SU(N) \times SU(24 -N) \label{eq:24-two}\\
b_1 = b_2 & = &  1 \nonumber\\
{\rm matter} & = &
3 ({\tiny\yng(1,1)}\times \cdot) +
3 (\cdot\times {\tiny\yng(1,1)}) +
1 ({\tiny\yng(1)}\times {\tiny\yng(1)}) \,.  \nonumber
\end{eqnarray}
Another family of models takes the form
\begin{eqnarray}
G & = &  SU(N) \times SU(12-N) \label{eq:two-s}\\
b_1 = b_2 & = &  2\nonumber\\
{\rm matter} & = &
6 ({\tiny\yng(1,1)}\times \cdot) +
6 (\cdot\times {\tiny\yng(1,1)}) +
4 ({\tiny\yng(1)}\times {\tiny\yng(1)}) \nonumber
\end{eqnarray}
for $N \leq 12$.
The family of models (\ref{eq:two-s}), including the single block
model with $b = 2, N = 12$ were previously constructed by
Schellekens using Gepner models \cite{Schellekens}.

There are a variety of other two-block combinations possible with
bifundamental matter and higher values of $b$'s.
When we consider larger values of $b_i, b_j,$ more interesting
combinations can also arise.  There are some models which contain
representations of the form ${\tiny\yng(1,1)}\times {\tiny\yng(1)}$.
For example, the two-block model with largest $N \leq M$
with such a representation has gauge group and matter content
\begin{eqnarray}
G & = &  SU(5) \times SU(7) \label{eq:two-5-7}\\
(b_1,  b_2) & = &  (4, 2)\nonumber\\
{\rm matter} & = &
2 ({\tiny\yng(1)}\times \cdot) +
1 ({\tiny\yng(1,1)}\times \cdot) +
3 ({\tiny\yng(2)}\times \cdot) +
2 (\cdot\times {\tiny\yng(1,1,1)}) +
2 ({\tiny\yng(1,1)}\times {\tiny\yng(1)}) +
2 ({\tiny\yng(1)}\times {\tiny\yng(1)})  \nonumber\\
H-V  & = & 273\,.\nonumber
\end{eqnarray}
There is a similar model with gauge group $SU(5) \times SU(6)$, but with $SU(5)$ adjoints instead of symmetric representations.
\begin{eqnarray}
G & = &  SU(5) \times SU(6) \label{eq:two-5-6}\\
(b_1,  b_2) & = &  (4, 2)\nonumber\\
{\rm matter} & = &
4 ({\tiny\yng(1)}\times \cdot) +
3 ({\rm Adj}\times \cdot) +
3 (\cdot\times {\tiny\yng(1,1,1)}) +
2 ({\tiny\yng(1,1)}\times {\tiny\yng(1)}) +
2 ({\tiny\yng(1)}\times {\tiny\yng(1)})  \nonumber\\
H-V  & = & 273\,.\nonumber
\end{eqnarray}
These models both saturate the gravitational anomaly, have similar
representation content, and satisfy the Kodaira constraint.

As the rank of the gauge group factors drops, more exotic matter
multiplets charged under two factors appear.  For example, for $SU(4)
\times SU(4)$ there are models containing matter which transforms in a
non-trivial non-fundamental representation of two gauge groups.  One
example is given by the model
\begin{eqnarray}
G & = &  SU(4) \times SU(4)\\
(b_1,  b_2) & = &  (2, 2) \nonumber\\
{\rm matter} & = &
32 ({\tiny\yng(1)}\, \times \cdot) +
32 (\cdot\times {\tiny\yng(1)}) +
1 ({\tiny\yng(1,1)}\times {\tiny\yng(1,1)}) \nonumber\\
H-V  & = & 262\,. \nonumber
\end{eqnarray}

Another interesting class of models are those which contain two blocks
$SU(N) \times SU(M)$ for large $M$ and small $N$. 
For example we find the following three  models
\begin{eqnarray}
G & = &  SU(2) \times SU(24)\\
(b_1,  b_2) & = &  (88, 1) \nonumber\\
{\rm matter} & = &
1({\tiny\yng(8)} \times \cdot) +
1({\tiny\yng(12)} \times \cdot)  
+1 ({\tiny\yng(2)}\times{\tiny\yng(1,1)}) \nonumber\\
H-V  & = & 272\,. \nonumber
\end{eqnarray}
\begin{eqnarray}
G & = &  SU(3) \times SU(24)\\
(b_1,  b_2) & = &  (22, 1) \nonumber\\
{\rm matter} & = &
1({\tiny\yng(6)} \times \cdot)  
+1 ({\tiny\yng(1)} \times {\tiny\yng(1,1)}) \nonumber\\
H-V  & = & 273\,. \nonumber
\end{eqnarray}
\begin{eqnarray}
G & = &  SU(2) \times SU(19)\\
(b_1,  b_2) & = &  (27,1) \nonumber\\
{\rm matter} & = &
1({\tiny\yng(1)} \times \cdot)
+2({\tiny\yng(2)} \times \cdot) +
1({\tiny\yng(3)}\times\cdot)  +
1({\tiny\yng(4)}\times\cdot)  \nonumber\\
&&+1({\tiny\yng(5)}\times\cdot)  +
1({\tiny\yng(6)}\times\cdot)  \nonumber\\
&&+1 ({\tiny\yng(1)}\times {\tiny\yng(1,1)})
+ 1 ({\tiny\yng(3)}\times {\tiny\yng(1)}) \nonumber\\
H-V  & = & 273\,. \nonumber
\end{eqnarray}
These are the only multiblock models
with a gauge group larger than $SU(18)$ that has non-bifundamental jointly
charged matter.  These models all severely violate the Kodaira bound.
It is perhaps interesting to note that models containing
$SU(N)$  factors
with $N=20, 21, 22, 23$ cannot have jointly charged matter other than
bifundamental matter as in the family of models (\ref{eq:24-two})

\subsection{Matter transforming under more than two factors}

We have also considered models containing more than two blocks which
when taken together satisfy the gravitational anomaly bound, and which
contain matter charged under more than two gauge group factors.  A
limited class of such multiply-charged matter representations are
known to appear in F-theory constructions.  In particular
tri-fundamental representations of $SU(2) \times SU(2) \times SU(N)$
can arise at a point where the singularity structure is enhanced to
$D_{N + 2}$ \cite{Katz-Vafa}, and tri-fundamentals of $SU(2) \times
SU(3) \times SU(N)$ can be realized from $E_{N + 3}$ singularities for
$N \leq 5$.  In \cite{KMT} we identified apparently consistent
low-energy models with $T = 1$ containing tri-fundamental matter
charged under the three gauge group factors $SU(2) \times SU(3) \times
SU(6)$.  While we have not done a completely systematic search, we
have identified a number of the interesting matter structures of this
type which can arise in $T = 0$ models.  We list here some of the
possibilities.  While this list is not necessarily comprehensive, it
should serve to demonstrate the kinds
of multiply-charged matter representations which may be possible.
\vspace*{0.05in}

\noindent
{\bf 3-charged matter}

As for $T = 1$, at $T = 0$
we find tri-fundamental matter charged under $SU(2)
\times SU(3) \times SU(6)$.
Such matter appears in the following 3-block model

\begin{eqnarray}
G & = &  SU(2)
\times SU(3) \times SU(6)\label{eq:236}\\
(b_1,  b_2, b_3) & = &  (3, 2, 1)\nonumber\\
{\rm matter} & = &
1 ({\tiny\yng(1)}\times 
{\tiny\yng(1)}\times 
{\tiny\yng(1)}) +
36 ({\tiny\yng(1)}\times 
\cdot\times 
\cdot) +
30 ( \cdot\times 
{\tiny\yng(1)}\times 
\cdot) +
12 ( \cdot\times 
\cdot\times 
{\tiny\yng(1)}) \nonumber \\
& &\hspace*{0.1in} +
1 ({\tiny\yng(2)}\times 
\cdot\times 
\cdot) +
3 (\cdot\times 
\cdot\times 
{\tiny\yng(1,1)}) 
\nonumber\\
H-V  & = & 272\,.\nonumber
\end{eqnarray}

Matter charged under $SU(2) \times SU(4) \times SU(4)$ appears in the
model\footnote{Thanks to D.\ Morrison for suggesting this possibility.}
\begin{eqnarray}
G & = &  SU(2)
\times SU(4) \times SU(4)\label{eq:244}\\
(b_1,  b_2, b_3) & = &  (2, 4, 4)\nonumber\\
{\rm matter} & = &
2({\tiny\yng(1)} \times{\tiny\yng(1)} \times{\tiny\yng(1)})
+4(\cdot \times{\tiny\yng(1)} \times{\tiny\yng(1)})
+2(\cdot \times {\tiny\yng(1,1)} \times {\tiny\yng(1,1)})
\nonumber \\
& &\hspace*{0.1in} 
+8({\tiny\yng(1)}\times \cdot \times \cdot)
+3(\cdot \times {\rm Adj} \times \cdot)
+3(\cdot \times \cdot \times {\rm Adj})
\nonumber\\
H-V  & = & 273\,.\nonumber
\end{eqnarray}

There is also matter charged under $SU(3) \times SU(3) \times SU(3)$,
appearing in the model
\begin{eqnarray}
G & = &  SU(3)
\times SU(3) \times SU(3)\label{eq:3-three}\\
(b_1,  b_2, b_3) & = &  (2, 2, 2)\nonumber\\
{\rm matter} & = &
1 ({\tiny\yng(1)}\times 
{\tiny\yng(1)}\times 
{\tiny\yng(1)}) +
1 \left[({\tiny\yng(1)}\times 
{\tiny\yng(1)}\times 
\cdot) + {\rm cyclic} \right] +
27 \left[ ({\tiny\yng(1)}\times 
\cdot\times 
\cdot) + {\rm cyclic} \right] 
\nonumber\\
H-V  & = & 273\,.\nonumber
\end{eqnarray}
Both these models containing tri-fundamental matter 
satisfy the Kodaira constraint.  There is also an interesting
combination of 3 blocks of the form (\ref{eq:4-box}) which contains
matter charged under $SU(4) \times SU(4) \times SU(4)$.

\begin{eqnarray}
G & = &  SU(4)
\times SU(4) \times SU(4)\label{eq:4-three}\\
(b_1,  b_2, b_3) & = &  (4, 4, 4)\nonumber\\
{\rm matter} & = &
4 ({\tiny\yng(1)}\times 
{\tiny\yng(1)}\times 
{\tiny\yng(1)}) +
1 \left[ ({\tiny\yng(2,2)}\times 
\cdot\times 
\cdot) + {\rm cyclic} \right] 
\nonumber\\
H-V  & = & 271\,.\nonumber
\end{eqnarray}

It is possible to combine four SU(3) blocks to have multiple
tri-fundamentals between groups of 3 of the SU(3)'s
\begin{eqnarray}
G & = &  SU(3)
\times SU(3) \times SU(3) \times SU(3)\label{eq:3-four}\\
(b_1,  b_2, b_3, b_4) & = &  (3, 3, 3, 3)\nonumber\\
{\rm matter} & = &
1 \left[ ({\tiny\yng(1)}\times 
{\tiny\yng(1)}\times 
{\tiny\yng(1)}\times 
\cdot) + {\rm cyclic} \right] +
3 \left[({\tiny\yng(1)}\times 
{\tiny\yng(1)}\times  \cdot \times
\cdot) + {\rm 5\ permutations} \right]
\nonumber\\& &\hspace*{0.1in}
+
1 \left[ ({\tiny\yng(2,1)}\times 
\cdot\times 
\cdot\times\cdot) + {\rm cyclic} \right] 
\nonumber\\
H-V  & = & 270\,.\nonumber
\end{eqnarray}

\vspace*{0.05in}

\noindent
{\bf Matter charged under more than three factors}

We have found a few exotic models in which matter can be charged under
more than 3 gauge group factors.  

There is a combination of 4 SU(2) factors carrying a 4-fundamental in
a model which satisfies the Kodaira constraint

\begin{eqnarray}
G & = &  SU(2)^4\label{eq:2-5}\\
b_i & = & 4\nonumber\\
{\rm matter} & = &
2({\tiny\yng(1)}\times 
{\tiny\yng(1)}\times 
{\tiny\yng(1)}\times 
{\tiny\yng(1)}) +
8 \left[({\tiny\yng(1)}\times 
{\tiny\yng(1)}\times
  \cdot \times
\cdot) +  \text{5 permutations} \right] \nonumber\\
& &\hspace*{0.1in}
+
3 \left[ ({\tiny\yng(2)}\times 
\cdot\times 
\cdot\times 
\cdot) + {\rm cyclic} \right] 
\nonumber\\
H-V  & = & 248\,.\nonumber
\end{eqnarray}

And there is a more exotic combination of 8 SU(2) factors at $b = 8$
where each block has 128 fundamental representations and one S4
(5-dimensional) representation

\begin{eqnarray}
G & = &  SU(2)^8\label{eq:2-8}\\
b_i & = & 8\nonumber\\
{\rm matter} & = &
1 ({\tiny\yng(1)}\times 
{\tiny\yng(1)}\times 
{\tiny\yng(1)}\times 
{\tiny\yng(1)}\times 
{\tiny\yng(1)}\times 
{\tiny\yng(1)}\times 
{\tiny\yng(1)}\times 
{\tiny\yng(1)}) \nonumber\\
& &\hspace*{0.1in} +
1 \left[ ({\tiny\yng(4)}\times 
\cdot\times 
\cdot\times 
\cdot\times 
\cdot\times 
\cdot\times 
\cdot\times 
\cdot) + {\rm cyclic} \right] 
\nonumber\\
H-V  & = & 272\,.\nonumber
\end{eqnarray}

\section{Conclusions and outlook}
\label{sec:conclusions}

\subsection{Summary of results}

We have systematically investigated the possible gauge groups and
matter structure for 6D ${\cal N} = 1$ supergravity models without
tensor multiplets.  We have restricted attention to gauge groups which
are a product of $SU(N)$ factors, although the methods used here would
apply equally well to general semi-simple gauge groups.  Anomaly
cancellation conditions provide strong constraints which limit the
range of possibilities for such models.   Potentially
interesting results of our analysis include the following:
\vspace*{0.05in}

\noindent
{\bf New matter representations:}

We have identified a number of $SU(N)$ matter representations which
are not ruled out by low-energy consistency conditions, but whose
realization in string theory is not yet known.  In F-theory, $T = 0$
models constitute the simplest class of compactifications on an
elliptically-fibered Calabi-Yau threefold, with a base manifold $B
=\P^2$.  Some of the
novel matter representations we have found are compatible with
topological F-theory constraints, and may be realized by new
codimension two singularities in F-theory.  Explicitly identifying and
classifying such codimension two singularities and associated matter
representations in F-theory will be addressed elsewhere.  It may also
be interesting to look for realizations of these matter
representations in other string vacuum constructions.

\noindent
{\bf Violation of Kodaira bound:}

All F-theory realizations of $T = 0$ 6D theories satisfy the Kodaira
constraint, which follows from the condition that the elliptic
fibration in the F-theory model must have a total space which is
Calabi-Yau.  The Kodaira constraint can be expressed in terms of a
condition on the field content and anomaly structure of the low-energy
6D theory.
There are a finite number of models, of which we have
described several explicitly here, which satisfy the anomaly
cancellation condition on the low-energy theory but which seem to
violate the Kodaira constraint.  It would be very interesting to know
whether these represent models which can be built using an as-yet
unknown string construction, or whether they suffer from some UV
inconsistency which may be identifiable from the structure of the
low-energy theory.  It is also possible that the map from
\cite{tensors} which we have used to pull back the Kodaira constraint
to the low-energy theory may need to be modified in the presence of
exotic singularities; this might make it possible to reconcile some of
the  models found here with F-theory constraints.
\vspace*{0.05in}

\noindent
{\bf Models with rigid symmetry:}

Many of the most unusual matter representations we have found live in
models which either completely or almost completely saturate the
gravitational anomaly bound $H-V \leq 273$.  When this bound is
saturated, there are no uncharged hypermultiplets, and any deformation
of the model will break the symmetry and reduce the matter content.
Thus, these models are delicately balanced configurations which exist
only at specific points in the moduli space of 6D supergravity
theories.  Many of the
explicit models we have found which go outside the domain of
established F-theory constructions turn out to precisely saturate the
gravitational anomaly bound and
exhibit remarkable
numerological/group theory structure, suggesting that some novel
stringy mechanism may enable the existence of these theories as
quantum-complete theories of supergravity.

\vspace*{0.05in}

\noindent
{\bf Diversity at low rank:}

When the rank of the gauge group factors is large, in general we find
that the associated models contain matter associated with well-known
F-theory singularity types and clearly satisfy the Kodaira constraint.
As the rank of the factors decreases, however, more exotic types of
matter appear and more models arise which violate the Kodaira
constraint.  Models containing only $SU(2)$ factors become difficult
to classify, and admit a wide range of representations.  This
observation matches with recent work on $U(1)$ factors in 6D
supergravity models, to appear elsewhere, which shows that the range
of matter charged under abelian factors is less constrained than for
matter charged under nonabelian factors.

It would be interesting to understand whether there is some systematic
reason for the increase in diversity of models at low rank.

\subsection{Outlook}

There are a number of further directions in which one can hope to gain
further insight into the theories described here.  As mentioned above,
further analysis of codimension 2 singularities in F-theory will be
helpful in identifying which of the models found here have acceptable
F-theory constructions.  It has also recently been suggested that a
more general class of brane constructions in F-theory may give rise to
additional matter content from nonabelian structure in the Higgs
field of the 7-brane world-volume theory \cite{Clay-Vafa}.  It would
be interesting to see if such constructions could give rise to some of
the exotic matter representations encountered here for models with $T
= 0$.  

We have focused here on $SU(N)$ gauge group factors.  Similar methods
can be used to analyze models with other types of gauge groups.  For
gauge groups $SO(10)$ and $E_6$ it was shown in \cite{Dienes-mr,
  Dienes} that while some exotic representations of these groups can
be found in free-field heterotic constructions of four-dimensional
theories, other representations, such as the {\bf 126} of $SO(10)$, or
representations larger than the {\bf 78} of $E_6$, are impossible in
this type of model.  It would be interesting to compare such
constraints from heterotic constructions with the classification of
allowed six-dimensional theories with these gauge groups.  For
example, in 6D supergravity theories, many large representations of
$SO(10)$ and $E_6$ will be ruled out by the gravitational anomaly
bound on the number of hypermultiplets, and the possible matter
representations for these groups will be even further constrained by
the Kodaira condition.  We leave further exploration of this question
to further work.

Another point of view which may be helpful in understanding the models
described here is to consider connectivity of the space of theories.
By Higgsing some of the theories with higher symmetry and more
complicated matter content, the models reduce in complexity to models
with better understood stringy descriptions.  For example, by
successively Higgsing pairs of fundamental representations in the
$SU(8)$ model (\ref{eq:b2b}) with the exotic 4-antisymmetric
representation, this model reduces to a standard $SU(5)$ model with
only fundamental and antisymmetric matter content.  Finding a string
realization of simpler models and moving backwards up such a Higgsing
chain may help us to understand how the more complicated matter
structures found here can be realized in string theory.  Other
continuous transitions should be possible which connect the various
branches of the 6D $T = 0$ moduli space, for example by tuning
parameters which change the ``degree'' $b_i$ of the various $SU(N_i)$
factors.  Finally, there are points in the moduli space where
tensionless strings arise and 29 hypermultiplets are traded for a
single tensor multiplet.  Such transitions connect the space of models
described here to the $T = 1$ models described in \cite{KMT}.  Note,
however, that such transitions cannot occur for many of the most
exotic models found here since when the gravitational anomaly is
almost saturated there are not enough neutral hypermultiplets to
effect a transition of this kind.  Thus, some of the most exotic
phenomena we have identified here  are probably unique to models without
tensor multiplets.

We hope that the variety of new apparently consistent supergravity
models identified here will stimulate
some further understanding of new string constructions or will help to
generate new constraints on quantum theories of gravity.  Lessons of
this type for the moduli space of 6D supergravity theories may have
implications for our understanding of 4D gravity coupled to gauge
groups and matter.
\vspace*{0.2in}

\appendix
\noindent
{\large {\bf Appendices}}

\section{Global anomalies}
In this section we  prove that locally non-anomalous blocks
with gauge group $SU(2)$ and $SU(3)$ in $T = 0$ theories
are free of global
anomalies of the kind addressed in \cite{Witten-SU2}.
The problem with $SU(2)$ and $SU(3)$ charged chiral fermions
is that the fermion measure might obtain a phase
under global gauge transformations, which are gauge transformations
that are not homotopic to the identity. This happens for only for the
gauge groups $SU(2)$ and $SU(3)$ among the $SU$ groups in
six dimensions because $\pi_6 (SU(2))=\mathbb{Z}_{12}$,
$\pi_6 (SU(3))=\mathbb{Z}_6$ while $\pi_6$ is trivial for the other
$SU(N)$ gauge groups.

Let us first consider $SU(3)$ in six dimensions. Defining the global
gauge transformation that generates $\pi_6 (SU(3))=\mathbb{Z}_6$
as $g$, we need to determine the phase $2 \pi \alpha_r$ a chiral fermion measure in
representation $r$ acquires when acted on by $g$.
Note that $\alpha_r$ is defined up to integers.

This problem was essentially solved in \cite{BV-global}, but let us
phrase it in a language convenient for our purposes.
The result of \cite{BV-global}, \cite{Elitzur} is that if an $SU(4)$
representation $R$ is broken into $\sum_i r_i$ of $SU(3)$
representations $r_i$ in a canonical embedding then
\be
\sum \alpha_{r_i} = {B_R \ov 3!}
\ee
Without loss of generality, let us embed the $SU(3)$ in the
upper left corner of $SU(4)$. 
Recall from section \ref{sec:results},
\begin{align}
B_R + 2C_R &= {1 \ov 2} \tr_R T_{12}^4 \\
C_R &= {3 \ov 4} \tr_R T_{12}^4 T_{34}^4
\end{align}
Note that $C_R$ is a multiple of $3$ as $\tr_R T_{12}^4 T_{34}^4 $
is integral, and is a multiple of $4$. This can be seen from looking at
which Young tableaux contribute to the trace for a given
representation (for an explicit proof, see \cite{tensors}).
If $R$ is broken into $\sum_i r_i$, it is clear that
\be
B_R + 2C_R = {1 \ov 2} \tr_R T_{12}^4=
\sum_i {1 \ov 2} \tr_{r_i} T_{12}^4=
2 \sum_i C_{r_i}
\ee
and therefore
\be
2 \sum_i C_{r_i}  \equiv  B_R \mod 6  \equiv  \sum_i 6 \alpha_{r_i} \mod 6
\ee
Since the measure of a chiral fermion in the trivial representation
does not acquire a phase under global transformations, and by
taking $R=\bold{4}$ we find that $\alpha_{\bold{3}} = {C_{\bold{3}}/ 3}$. It is
straightforward to carry this through by going through the representations
of $SU(4)$, and showing by induction that actually
\be
\alpha_{r} = {C_r \ov  3}
\ee

For $SU(2)$ the situation is similar. 
Denoting the generator of
$\pi_6 (SU(2))=\mathbb{Z}_{12}$  by $g'$ we need to find the phase
$2 \pi \alpha_{r'}$ the chiral fermion measure in representation
$r'$ of $SU(2)$ acquires when acted on by $g'$.

Let us embed $SU(2)$ into $SU(3)$. It is known(for example as stated in
\cite{BV-global}), that $g'$ maps to $g$ when we do the embedding.
Therefore we see that if an $SU(3)$ representation $r$ is broken
into $\sum_i r'_i$ of $SU(2)$ representations $r'_i$ in a canonical
embedding
\be
\sum_i \alpha_{r'_i} = \alpha_r = {C_r \ov 3} =\sum_i {C_{r'_i} \ov 3}
\ee
By an almost word-by-word duplication of the argument for $SU(3)$,
we obtain
\be
\alpha_{r'} = {C_{r'} \ov  3}
\ee

This is a satisfying result, because we see from the anomaly equations
on the $C$ factors
\be
\sum_i x_r \alpha_r-{\alpha_\text{adj}} = \sum_i x_r {C_r \ov 3} -{C_\text{adj}\ov 3 }=b^2
\ee
which is an integer when $b$ is an integer.
Note that the far left hand side is the phase (divided by $2 \pi$)
the fermion measure obtains under the global transformation
given by the generator of $\pi_6$ of the gauge group.
Therefore if the far right hand side is integral our theory would
not have any global gauge anomalies.
But as proven in \cite{tensors}, for $T=0$, $b$ is always integral
for non-anomalous blocks. So an $SU(2)$ or $SU(3)$ block of a $T = 0$
theory free of perturbative anomalies does not have global anomalies!

\section{Proof of bounds on $b$}

Using group theory identities, one can find constraints
on the matter content of individual blocks. In this Appendix,
we  show that even without the $\nhv$ constraint
we can bound the degree
$b$ of an $SU(N)$ block just from group theory constraints.
The only equations we  use are the anomaly cancellation conditions
\begin{align}18 b_i+A^i_\text{adj}&=
\sum_R x^i_R A^i_R \label{aeq}\\
B^i_\text{adj}&=\sum_R x^i_R B^i_R \label{beq}\\
3b_i^2+C^i_\text{adj} &=
\sum_R x^i_R C^i_R \label{ceq}
\end{align}

In section \ref{ss:weyl} we  make some useful
statements based on the Weyl character formula.
In section \ref{ss:conseq} we  see how this bounds $b$
for gauge groups larger than $SU(3)$.
In section \ref{ss:su2su3} we  discuss the process of
bounding $b$'s for the gauge groups $SU(2)$ and $SU(3)$.
In section \ref{ss:gthysum}
we  summarize the results.
A useful reference for this section is
\cite{Cahn:1985wk}, chapters 12 and 13.

\subsection{The Weyl Character Formula} \label{ss:weyl}
We will use the Weyl character formula (equation (XIII.37) of
\cite{Cahn:1985wk})
\be
\tr_\lambda e^{\rho} =
{\sum_{w \in W} \text{sign}(w)  e^{\langle\lambda+\delta,w \rho\rangle} \ov
{\sum_{w \in W} \text{sign}(w)  e^{\langle\delta,w \rho\rangle}}}
\ee
In this formula $\tr_\lambda$ denotes the trace of the representation
with highest weight vector $\lambda$. 
$\rho$
is an element of the Lie algebra $\rho=\rho_a T^a$ where $T^a$
is the Cartan basis, and $(\rho_1 , \cdots, \rho_r)$
are coordinates on the weight space.
Brackets denote inner products in the weight space.
$R^+$ is the set of positive roots of the Lie algebra, and
$W$ is the Weyl group.
The vector $\delta$ is defined to be half the sum of the positive roots
\be
\delta = {1 \ov 2} \sum_{\alpha \in R^+} \alpha
\ee

For $\rho =s \delta$ the equation simplifies to
\begin{align}
\begin{split}
&\tr_\lambda e^{s\delta} \\
&= \prod_{\alpha \in R^+} 
\left( {e^{{s \ov 2}\langle\alpha,\lambda+\delta\rangle}-e^{-{s \ov 2}\langle\alpha,\lambda+\delta\rangle} \ov
e^{{s \ov 2}\langle\alpha,\delta\rangle}-e^{-{s \ov 2}\langle\alpha,\delta\rangle}} \right) \\
&= \prod_{\alpha \in R^+}  { \langle\alpha,\lambda+\delta\rangle \ov\langle\alpha,\delta\rangle } 
\prod_{\alpha \in R^+}
\left({ 1+ {1 \ov 6}  \langle\alpha,\lambda+\delta\rangle^2 ({s \ov 2})^2
+ {1 \ov 120}  \langle\alpha,\lambda+\delta\rangle^4 ({s \ov 2})^4 +\cdots \ov
1+ {1 \ov 6}  \langle\alpha,\delta\rangle^2 ({s \ov 2})^2 +
{1 \ov 120}  \langle\alpha,\delta\rangle^4 ({s \ov 2})^4 + \cdots} \right)
\end{split}
\end{align}
This is due to the relation (equation (XIII.13) of \cite{Cahn:1985wk})
\be
\sum_{w \in W} \text{sign}(w)  e^{\langle\delta,w \rho\rangle}
=\prod_{\alpha \in R^+} ( e^{{1 \ov 2}\langle\alpha,\rho\rangle}-e^{-{1 \ov 2}\langle\alpha,\rho\rangle} )
\ee
which holds for an arbitrary vector $\rho$.

Expanding in $s$, and looking at the terms of order 0, 2, and 4, we find that
\begin{align}
\label{ABCD}
\begin{split}
D_\lambda &= \tr_\lambda 1 =  \prod_{\alpha \in R^+}
 { \langle\alpha,\lambda+\delta\rangle \ov\langle\alpha,\delta\rangle }  \\
A_\lambda (\tr_f \delta^2) &=  \tr_\lambda \delta^2
= {D_\lambda \ov 12} \sum_{\alpha \in R^+} (\langle\alpha,\lambda+\delta\rangle^2 - \langle\alpha,\delta\rangle^2) \\
B_\lambda (\tr_f \delta^4) + C_\lambda (\tr_f \delta^2)^2 &= 
\tr_\lambda \delta^4 =
{D_\lambda \ov 120} \sum_{\alpha \in R^+} (-\langle\alpha,\lambda+\delta\rangle^4 + \langle\alpha,\delta\rangle^4) \\
&\qquad \qquad + {D_\lambda \ov 48} (\sum_{\alpha \in R^+} (\langle\alpha,\lambda+\delta\rangle^2 - \langle\alpha,\delta\rangle^2))^2
\end{split}
\end{align}

Now $\tr_f \delta^2$ and $\tr_f \delta^4$ are explicitly calculable.
We consider $SU(N)$ groups with the normalization
$\tr_f T^a T^b =2 \delta_{ab}$.
We use the fact that for $SU(N)$ the positive roots are given by
\be
\alpha_{ij} =\alpha_i+\alpha_{i+1} + \cdots + \alpha_{j-1} + \alpha_j
\ee
for $i \leq j$ where $\alpha_i$ for $i=1,2,\cdots,N-1$
are the simple roots of $SU(N)$ whose Cartan
matrix is given by
\be
(\alpha_i \cdot \alpha_j ) =
\begin{pmatrix}
2&-1&0&\cdots&0&0\\
-1&2&-1&\cdots&0&0\\
0&-1&2&\cdots&0&0\\
\vdots&\vdots&\vdots&\ddots&\vdots&\vdots\\
0&0&0&\cdots&2&-1\\
0&0&0&\cdots&-1&2
\end{pmatrix}
\ee
Then,
\be
\delta= {1 \ov 2} \sum_{i \leq j} \alpha_{ij}
= \sum_{i=1}^{N-1} {i(N-i) \ov 2} \alpha_i
\ee

By explicit calculation one may show that
\be
\langle \alpha_i , \delta \rangle = 1
\ee
and therefore taking the dual basis of $\{ \alpha_i \}$
to be $\{ \beta_i \}$,
\be
\delta = \sum_i \beta_i
\ee

Now we use the fact that the highest weight vector for the
fundamental representation is $\beta_1$.
Then using \eq{ABCD} we find that
\begin{align}
\begin{split}
\tr_f \delta^2
&= {N \ov 12} \sum_{i \leq j} (\langle\alpha_{ij},\beta_1+\delta\rangle^2 - \langle\alpha_{ij},\delta\rangle^2) \\
&= {N \ov 12} \sum_{i \leq j} ((\delta_{i1} +(j-i+1))^2 - (j-i+1)^2) \\
&= {N \ov 12} \sum_{j=1}^{N-1} ((j+1)^2 - j^2) = {N(N-1)(N+1) \ov 12}
\end{split}
\end{align}
and likewise
\begin{align}
\begin{split}
\tr_f \delta^4
&= {N \ov 120} \sum_{i \leq j}  (-\langle\alpha_{ij},\beta_1+\delta\rangle^4 + \langle\alpha_{ij},\delta\rangle^4)
+ {N \ov 48} \sum_{i \leq j}  (\langle\alpha_{ij},\beta_1+\delta\rangle^2 - \langle\alpha_{ij},\delta\rangle^2))^2 \\
&= {N \ov 120} \sum_{i \leq j}  (-(\delta_{i1} +(j-i+1))^4 +(j-i+1)^4)
+ {N \ov 48} (N^2-1)^2 \\
&= {N \ov 120} \sum_{j=1}^{N-1}  (-(j+1)^4 +j^4)
+ {N \ov 48} (N^2-1)^2 \\
&=- {N \ov 120}(N^4-1) + {N \ov 48} (N^2-1)^2 \\
&={1 \ov 240} N(N-1)(N+1)(3N^2 -7)
\end{split}
\end{align}

Furthermore, plugging in the equation for $A_\lambda$ to the equation
for the fourth order invariants and dividing both sides by
$(\tr_f \delta^2)^2$ we find that
\begin{align}
\label{ineq3}
\begin{split}
y_N B_\lambda  + C_\lambda
\leq {3A_\lambda^2 \ov  D_\lambda}
\end{split}
\end{align}
where we have defined
\be
y_N \equiv {\tr_f \delta^4 \ov (\tr_f \delta^2)^2} = {3 (3N^2-7)\ov 5 N(N^2-1)}
\ee

All the results of the current section hold for $SU(2)$ and $SU(3)$ also
if we set $B_\lambda =0$ by hand, which we can certainly do for these groups.

\subsection{Restriction on $b$} \label{ss:conseq}
For a single $SU(N)$ block with $N \geq 4$ define
\be
{\sum_R x_R (y_N B_R + C_R) \ov \sum_R x_R A_R}
= {3b^2 + 6+2N y_N  \ov 18b + 2N} \equiv \eta
\ee
This means that there must exist a representation $R_0$ with
\be
{y_N B_{R_0} + C_{R_0}  \ov A_{R_0} } \geq \eta
\ee
since by definition, $y_N B_{R_0} + C_{R_0}$ and
$A_{R_0}$ are positive. Then by inequality \eq{ineq3},
\be
({D_{R_0} \ov 3}) \eta \leq A_{R_0} \leq \sum_R x_R A_R = 18b+2N
\ee
Plugging in the definition of $\eta$, we find that
\be
D_{R_0} \leq {324 (b+N/9)^2 \ov b^2 + (2+2y_N N /3)}
\ee
The maximum value for the right hand side of the above equation
is obtained for $b=(18/N+6y_N)$ and plugging in this value of $b$
we obtain the inequality
\be
D_{R_0} \leq 324+ {4N^2 \ov 2 + 2y_NN/3 } \equiv D_N
\ee
Hence we obtain
\be
{b^2 + 2+2N y_N/3  \ov 18b + 2N} =
{\eta \ov 3} \leq {\mathop{\text{max}}_{R: D_R < D_N}}
( {A_R \ov D_R})
\label{eq:b-bound}
\ee
This procedure gives an upper bound $b_u$ on $b$, for all $N \geq 4$.
For $N=2,3$ we are able to obtain slightly improved bounds as we have
explicit expressions for $A_R, C_R, D_R$ for these groups.
This is helpful since the enumeration of $SU(2)$ and $SU(3)$ blocks
takes much more time numerically compared to blocks with $N \geq 4$.
We will elaborate on this in section \ref{ss:su2su3}.

Implementing \eq{eq:b-bound} to obtain a upper bound of $b$
is a rather tedious, but straightforward task. In practice we must find all
representations with $D_R < D_N$ and find the maximum value of
$A_R/ D_R$ among those representations. This can be done by using
the following useful
\vspace*{0.1in}

\noindent
\textbf{Fact : } Given two representations $R_1$ and $R_2$ represented
by young-diagrams $Y_1$, $Y_2$, if $Y_2$ can be obtained by attaching
columns of blocks to the right of $Y_1$, then necessarily $D_{R_1} < D_{R_2}$.
\vspace*{0.1in}

This follows simply from the fact that the dimension of a
representation is associated with the number of distinct labelings of
the boxes which are horizontally non-decreasing and vertically
increasing.  Adding columns to the right, there is always at least one
labeling of $Y_2$ for each labeling of $Y_1$ by simply repeating
entries on each row in the added columns.
\noindent
For example, if we define
\be
Y_1 = {\tiny\yng(4,4,4,3,2)}, \qquad Y_2 = {\tiny\yng(5,5,4,3,2)},
\qquad Y_3={\tiny\yng(5,5,4,4,2)}
\ee
the dimension of representation $Y_1$ is smaller than that of $Y_2$.
Meanwhile, it is not guaranteed that the dimension of $Y_1$ is smaller
than the dimension of $Y_3$.

Starting from single column representations one may span a tree of young
diagrams by attaching columns of varying length to the right until one runs into
a diagram with $D_R >D_N$. Since the dimension strictly increases at each step,
all the branches of the tree will eventually terminate and one will be able to
obtain all the representations with bounded dimension.

Although we have thus found an upper bound on $b$ for each group
$SU(N)$, which in principle makes the problem of enumerating blocks
into a finite algorithm, in practice it is helpful to reduce the bound
somewhat to make the enumeration of blocks more tractable.
It turns out that we can further restrict $b$ by utilizing the condition
\be
A_{R_0} \leq 18b_u+2N
\ee
That is, it can be the case that
\be
{b^2 + 2+2N y_N/3  \ov 18b + 2N} = {\eta \ov 3} \leq
\mathop{\mathop{\text{max}}_{R: D_R < D_N }}_{\text{and} ~A_R \leq18b_u+2N} ( {A_R \ov D_R})
\ee
further restricts $b$ below the bound coming from (\ref{eq:b-bound}).

For example, in the case of $SU(7)$ one finds that
\be
{\mathop{\text{max}}_{R: D_R < 386}}( {A_R \ov D_R}) = {495 \ov 492} = 1.07 \cdots
\ee
and hence
\be
b \leq 19
\ee
But one finds that for $b\leq 19$, $18b+2N \leq 356$, so
\be
\mathop{\mathop{\text{max}}_{R: D_R < 386}}_{\text{and}~A_R \leq 356}
( {A_R \ov D_R}) = {165 \ov 210} = 0.78 \cdots
\ee
and hence $b$ is further restricted to the value
\be
b \leq 14
\ee
The bound on $b$ obtained this way for $4 \leq N \leq 17$ is given in table \ref{t:bbound}.
\begin{table}[t!]
\centering
\setlength{\extrarowheight}{2pt}
  \begin{tabular}{|c|c|c|c|c|c|c|c|c|c|c|c|c|c|c|} 
  \hline
  $N$ & 3& 4 & 5 & 6 & 7 & 8 & 9 & 10 & 11  & 12 & 13 & 14& 15 & $N \geq 16$\\ \hline
 Bound on $b$  & 617& 126 & 40 & 24 & 14 & 12 & 11 & 6& 6&5 &3 & 3 &3 & 2\\ \hline
  \end{tabular}
  \caption{Upper bound on $b$ for individual block of group $SU(N)$.}
\label{t:bbound}
\end{table}
For $N \geq 18$, 
\be
324+ {4N^2  \ov 2 + 2y_N N/3 } < {N(N-1)(N-2) \ov 6}
\ee
This means that the representation with
maximum $A/D$ in an $SU(N)$, $N \geq 18$ block is
either the adjoint, symmetric, antisymmetric or fundamental.
The maximum value of $A/D$ of these is given by
the symmetric and hence it must be the case that
\be
{b^2 + 2+2N y_N/3  \ov 18b + 2N}  \leq {2(N+2) \ov N(N+1)}
\ee
A little bit of algebra shows that this implies $b \leq 2$
for $N \geq 18$. 
This completes the data needed for Table \ref{t:bbound}.
Given the upper bound on $b$ for each $N$ it is therefore a finite
problem to enumerate all possible gauge factor + matter ``blocks''.
As noted in section \ref{sec:results},
in most cases the actual maximum $b$ for each $N$ is smaller than that
given in Table \ref{t:bbound}.
In particular, above $N = 12$ only blocks with $b = 1$ are possible.

\subsection{Comments on $SU(2)$ and $SU(3)$ Blocks} \label{ss:su2su3} 

The $b$ values of the $SU(2)$  blocks
can be restricted in an equivalent fashion. The reason we
are addressing them separately is because the bound
on $b$ obtained for $SU(2)$  using the equations
in the previous section is very high. In particular,
the most naive bounds on $b$ for $SU(2)$
is of order $10^5$.

We will first provide the most naive bounds on $b$ we can
get for $SU(2)$ and $SU(3)$.
As mentioned in the previous section we can get slightly better
bounds because the equations for $A, C, D$ are simple enough
to manipulate directly.

For $SU(2)$ all representations are $m$-symmetric representations.
The dimension and group theory coefficients of the representation are
\begin{eqnarray}
A_m & = &  m (m +1) (m + 2)/6\\
C_m & = & A_m (3m^2 + 6m-4)/10 \\
D_m & = &  m +1\,.
\end{eqnarray}
Also recall that $B_m=0$ for all representations.
The anomaly equations \eq{aeq}, \eq{ceq} can be written as,
\begin{eqnarray}
\sum_m m (m +1) (m + 2) x_m & = &108 b + 24\\
\sum_m m^2 (m+2)^2 (m+1) x_m & = & 60b^2 + 36b + 192 \,.
\end{eqnarray}
Taking the largest $m$ with $x_m \neq 0$ to be $M$, we find that,
\be
{10b^2 + 6b + 32 \ov 18 b + 4} \leq M(M+2)
\ee
and hence,
\be
\left( {10b^2 + 6b + 32 \ov 18 b + 4} \right)^{3/2}
\leq M(M+2) \sqrt{M(M+2)}
< M(M+2) (M+1)
\leq 108b+24
\ee
This gives the bound $b \leq 68018$.

The representations of $SU(3)$ can be represented
by a pair of numbers $x$ and $y$ which denote the
number of two block/one block columns of its young diagram.
Then the dimension and group theory coefficients of
representation $(x,y)$ are given by,
\begin{eqnarray}
A_{x,y} & = &  {1 \ov 24}XY(X+Y)(X^2 + Y^2 + XY-3)\\
C_{x,y} & = &  {1 \ov 120}XY(X+Y)(X^2 + Y^2 + XY-3)(X^2 + Y^2 + XY-{9 \ov 2}) \\
D_{x,y} & = &  {1 \ov 2}{XY(X+Y)}\,.
\end{eqnarray}
where we have defined $X = (x+1)$, $Y=(y+1)$.
Writing out the anomaly equations and going through a
similar process as in the $SU(2)$ case one obtains the bound
$b \leq 617$.

Up to now we have been using that fact that in order for a block
to satisfy the anomaly equations, we must have some
representation $R$ with a large
\be
{C_R \ov A_R} \geq {3b^2 + C_\text{adj} \ov 18b + A_\text{adj}} \sim \mathcal{O}(b)
\ee
We have been ruling out $b$ values for which all such large
representations $R$ have
$A_R > 18b+A_\text{adj}$.
Finding all solutions to the anomaly equations
for $SU(2)$ and $SU(3)$ for the range of $b$
values constrained only by this condition
turns out to be a demanding task numerically,
and it is useful to further restrict the allowed values of $b$.
To do this, we generalize the strategy employed up to now.

Suppose $b_0 $ is a value not ruled out by the previous arguments.
This means that there exists a representation $R$ satisfying
\be
{C_R \ov A_R} \geq {3b_0 ^2 + C_\text{adj} \ov 18b_0 + A_\text{adj}}
\ee
and
\be
A_R \leq 18b_0 +A_\text{adj}
\ee
within the block.
Let $S(b_0)=\{ R_1 , \cdots, R_k \}$ be all the representations that
satisfy these two equations.
The fact that $A$, $C$, $D$, $A/D$ and $C/A$
are all strictly increasing functions with respect to $m$
in the case of $SU(2)$ and of $x$ and $y$ in the case of $SU(3)$
is helpful in constructing this list.

Now assume that we have the representation
$R_1$ in a block with given $b_0$.
Now if $R_1$ satisfied,
\be
3b_0 ^2 + C_\text{adj} = C_{R_1},
\qquad \text{and} \qquad
18b_0 +A_\text{adj}=A_{R_1}
\ee
we would have a solution for a single block whose
matter content is given by just one $R_1$.
Suppose this were not the case. \footnote{In fact, we can show that
for $SU(2)$ and $SU(3)$ there cannot be a block with a single
matter representation, i.e. a block with $\sum_R x_R =1$.}
By the same line of argument as before we must
have a representation $R$ satisfying,
\be
{C_R \ov A_R} >
{3b_0 ^2 + C_\text{adj}-C_{R_1} \ov 18b_0 + A_\text{adj}-A_{R_1}}
\ee
and
\be
A_R < 18b_0 +A_\text{adj}-A_{R_1}
\ee
If no such representation $R$ exists, the representation
$R_1$ cannot show up in a block with $b=b_0$.
If the situation were same for all the representations in
$S(b_0)=\{ R_1, R_2 , \cdots, R_k \}$ we could rule out
$b=b_0$. We can iterate this process to further rule out $b$ values.

We have employed this procedure for $SU(2)$ and the initial
bound $68,018$ has been pulled down to $36,647$.
For $SU(3)$ we have iterated this process 5 times and were
able to rule out $288$ values of $b$ in the range $b \leq 617$.

We can describe the problem of constructing single block models
in a very concise way as a partition problem for $SU(2)$ and $SU(3)$
as the coefficients of the anomaly equations are all positive for these
cases. While the situation is similar for $SU(3)$ we will for simplicity
only depict the process for $SU(2)$.
The problem is to find a combination of representations where
the $A_R, C_R, D_R$ values add up to $(4 + 18b, 8 + 3b^2, D)$
with $D \leq 276$. This is a straightforward partition problem, whose
algorithmic solution is simplified by the fact that $A_m/D_m$ and
$C_m/A_m$ are monotonically increasing functions of $m$.  We have
implemented an algorithm which computes all such partitions for fixed
$b$ and checked a representative sample of $b$'s in the allowed range
as described in the main text.

One last note is that the maximum $b$ value possible for single block
models turns out to be within an order of magnitude
of the upper bounds for small gauge groups. For $SU(3)$
the maximum $b$ possible for an $SU(3)$ is $b=597$ while the
bound is 617, and for $SU(2)$ we were able to find a block with
$b=24,297$ while the bound is given by $36,647$.

\subsection{Summary} \label{ss:gthysum}

To summarize, just from the group theory we find that each individual
block cannot have a gauge group larger than $24$.  The $b$ values are
bounded as in table \ref{t:bbound} $SU(N)$ with $N\geq 3$.  The best
bound we have for $SU(2)$ is $36,647$.  As discussed in the main text,
given an upper bound on $b$, for each fixed $N \leq 24$ we can solve
the finite partition problem for each $b$ to enumerate all possible
blocks.
\vspace*{0.1in}

\noindent
{\bf Acknowledgements}: We would like to thank  
Massimo Bianchi, David Morrison, Bert Schellekens, and
Cumrun Vafa
for helpful discussions.  
WT would like to thank the Institute for Physics and Mathematics of
the Universe (IPMU) for hospitality during part of this
work.
This research was
supported in part by the DOE under contract \#DE-FC02-94ER40818 and
in part by the National Science Foundation under Grant No. PHY05-51164.
DP also
acknowledges support as a String Vacuum Project Graduate Fellow,
  funded through NSF grant PHY/0917807.


\begin{thebibliography}{99}

\bibitem{variety}

The following papers give a representative sampling of the wide range
of 6D string vacuum constructions for theories with one or more tensor
multiplets (see also many of the other papers listed in the
references below, which are cited in more specific contexts in the
text):

M.~Bianchi, A.~Sagnotti,
  ``On the systematics of open string theories,''
  Phys.\ Lett.\  {\bf B247}, 517-524 (1990);
``Twist symmetry and open string Wilson lines,''
  Nucl.\ Phys.\  {\bf B361}, 519-538 (1991);

M.~J.~Duff, R.~Minasian and E.~Witten,
  ``Evidence for Heterotic/Heterotic Duality,''
  Nucl.\ Phys.\  B {\bf 465}, 413 (1996)
  {\tt arXiv:hep-th/9601036};

  E.~G.~Gimon and J.~Polchinski,
  ``Consistency Conditions for Orientifolds and D-Manifolds,''
  Phys.\ Rev.\  D {\bf 54}, 1667 (1996)
  {\tt arXiv:hep-th/9601038};

  A.~Dabholkar and J.~Park,
  ``An Orientifold of Type-IIB Theory on $K3$,''
  Nucl.\ Phys.\  B {\bf 472}, 207 (1996)
  {\tt arXiv:hep-th/9602030};
  ``Strings on Orientifolds,''
  Nucl.\ Phys.\  B {\bf 477}, 701 (1996)
  {\tt arXiv:hep-th/9604178};

G.~Aldazabal, A.~Font, L.~E.~Ibanez and F.~Quevedo,
  ``Heterotic/Heterotic Duality in D=6,4,''
  Phys.\ Lett.\  B {\bf 380}, 33 (1996)
  {\tt arXiv:hep-th/9602097}.

  E.~G.~Gimon and C.~V.~Johnson,
  ``K3 Orientifolds,''
  Nucl.\ Phys.\  B {\bf 477}, 715 (1996)
  {\tt arXiv:hep-th/9604129};

J.~Polchinski,
  ``Tensors from K3 orientifolds,''
  Phys.\ Rev.\  D {\bf 55}, 6423 (1997)
  {\tt arXiv:hep-th/9606165};


  J.~D.~Blum and A.~Zaffaroni,
  ``An orientifold from F theory,''
  Phys.\ Lett.\  B {\bf 387}, 71 (1996)
  {\tt arXiv:hep-th/9607019};

  P.~S.~Aspinwall,
  ``Point-like instantons and the spin(32)/Z(2) heterotic string,''
  Nucl.\ Phys.\  B {\bf 496}, 149 (1997)
  {\tt arXiv:hep-th/9612108};
  
  
  P.~Candelas, E.~Perevalov and G.~Rajesh,
   ``Toric geometry and enhanced gauge symmetry of F-theory/heterotic  vacua,''
  Nucl.\ Phys.\  B {\bf 507}, 445 (1997)
  {\tt arXiv:hep-th/9704097};
  ``Matter from toric geometry,''
  Nucl.\ Phys.\  B {\bf 519}, 225 (1998)
  {\tt arXiv:hep-th/9707049};

  P.~S.~Aspinwall, D.~R.~Morrison,
  ``Point - like instantons on K3 orbifolds,''
  Nucl.\ Phys.\  {\bf B503}, 533-564 (1997).
  {\tt arXiv:hep-th/9705104};
 

L.~E.~Ibanez and A.~M.~Uranga,
  ``D = 6, N = 1 string vacua and duality,''
  {\tt arXiv:hep-th/9707075};

  Z.~Kakushadze, G.~Shiu and S.~H.~H.~Tye,
  ``Type IIB orientifolds with NS-NS antisymmetric tensor backgrounds,''
  Phys.\ Rev.\  D {\bf 58}, 086001 (1998)
  {\tt arXiv:hep-th/9803141};

  C.~Angelantonj,
  ``Comments on open-string orbifolds with a non-vanishing B(ab),''
  Nucl.\ Phys.\  B {\bf 566}, 126 (2000)
  {\tt arXiv:hep-th/9908064};

 P.~S.~Aspinwall, S.~H.~Katz, D.~R.~Morrison,
  ``Lie groups, Calabi-Yau threefolds, and F theory,''
  Adv.\ Theor.\ Math.\ Phys.\  {\bf 4}, 95-126 (2000)
  {\tt hep-th/0002012};

  R.~Blumenhagen, V.~Braun, B.~Kors and D.~Lust,
  ``Orientifolds of K3 and Calabi-Yau manifolds with intersecting D-branes,''
  JHEP {\bf 0207}, 026 (2002)
  {\tt arXiv:hep-th/0206038};

  S.~Hellerman, J.~McGreevy and B.~Williams,
  ``Geometric Constructions of Nongeometric String Theories,''
  JHEP {\bf 0401}, 024 (2004)
  {\tt arXiv:hep-th/0208174};

G.\  Honecker,
``Massive U(1)s and heterotic five-branes on K3,''
Nucl. Phys. B 748 (2006) 126  {\tt arXiv:hep-th/0602101};

G.\  Honecker and M.\  Trapletti,
``Merging heterotic orbifolds and K3 compactifications with line bundles,''
JHEP 0701 (2007) 051  {\tt arXiv:hep-th/0612030}.

  


\bibitem{universality}
  V.~Kumar and W.~Taylor,
  ``String Universality in Six Dimensions,''
  {\tt arXiv:0906.0987 [hep-th]}.

\bibitem{finite}
  V.~Kumar and W.~Taylor,
  ``A bound on 6D ${\cal N} = 1$ supergravities,''
    JHEP {\bf 0912}, 050 (2009)
  {\tt arXiv:0910.1586 [hep-th]}.

\bibitem{KMT}
  V.~Kumar, D.~R.~Morrison and W.~Taylor,
  ``Mapping 6D ${\cal N} = 1$ supergravity to F-theory,''
  {\tt arXiv:0911.3393 [hep-th]}.

\bibitem{tensors}
  V.~Kumar, D.~R.~Morrison and W.~Taylor,
  ``Global aspects of the space of
 ${\cal N} = 1$ supergravities,''
  {\tt arXiv:1008.1062 [hep-th]}.

\bibitem{singer}
W.~Taylor,
  ``Anomaly constraints and string/F-theory geometry in 6D quantum
gravity,''
to appear in proceedings of {\sl ``Perspectives in Mathematics and
Physics,''} conference in honor of I.\ M.\  Singer,
  {\tt arXiv:1008.1062 [hep-th]}.


\bibitem{Seiberg-Witten}
  N.~Seiberg and E.~Witten,
  ``Comments on String Dynamics in Six Dimensions,''
  Nucl.\ Phys.\  B {\bf 471}, 121 (1996)
  {\tt arXiv:hep-th/9603003}.


\bibitem{Witten}
E.~Witten, ``Phase transitions in {$M$}-theory and {$F$}-theory,''  Nucl.
  Phys.  B {\bf 471} (1996) 195--216, {\tt arXiv:hep-th/9603150}.

\bibitem{Morrison-Vafa-II}
  D.~R.~Morrison and C.~Vafa,
  ``Compactifications of F-Theory on Calabi--Yau Threefolds -- II,''
  Nucl.\ Phys.\  B {\bf 476}, 437 (1996)
  {\tt arXiv:hep-th/9603161}.

\bibitem{Bianchi} 
  C.~Angelantonj, M.~Bianchi, G.~Pradisi, A.~Sagnotti and Y.~S.~Stanev,
  ``Comments on Gepner models and type I vacua in string theory,''
  Phys.\ Lett.\  B {\bf 387}, 743 (1996)
  {\tt arXiv:hep-th/9607229}.

\bibitem{Schellekens}
K.\ Schellekens, {\it unpublished}

\bibitem{Morrison-Taylor}
D.\ R.\ Morrison and W.\ Taylor, {\it to appear}.

\bibitem{Romans}
  L.~J.~Romans,
  ``Selfduality For Interacting Fields: Covariant Field Equations For
  Six-Dimensional Chiral Supergravities,''
  Nucl.\ Phys.\  B {\bf 276}, 71 (1986).



\bibitem{Sagnotti}
  A.~Sagnotti,
  ``A Note on the Green-Schwarz mechanism in open string theories,''
  Phys.\ Lett.\  B {\bf 294}, 196 (1992)
  {\tt arXiv:hep-th/9210127}.
  
\bibitem{Sadov}
  V.~Sadov,
  ``Generalized Green-Schwarz mechanism in F theory,''
  Phys.\ Lett.\  B {\bf 388}, 45 (1996)
  {\tt arXiv:hep-th/9606008}.

\bibitem{Green-Schwarz-West}
  M.~B.~Green, J.~H.~Schwarz and P.~C.~West,
 ``Anomaly Free Chiral Theories In Six-Dimensions,''
  Nucl.\ Phys.\  B {\bf 254}, 327 (1985).

\bibitem{Gates-Nishino}
S.\ J.\ Gates, Jr.\ and H.\ Nishino,
 ``Dual Versions of Higher Dimensional Supergravities and Anomaly Cancellations
in Lower Dimensions,''
  Nucl.\ Phys.\  B {\bf 268}, 532 (1986).

\bibitem{Witten-SU2}
  E.~Witten,
  ``An SU(2) anomaly,''
  Phys.\ Lett.\  B {\bf 117}, 324 (1982).

\bibitem{Vafa-f}
  C.~Vafa,
  ``Evidence for F-Theory,''
  Nucl.\ Phys.\  B {\bf 469}, 403 (1996)
  {\tt arXiv:hep-th/9602022}.
  
\bibitem{Morrison-Vafa-I}
  D.~R.~Morrison and C.~Vafa,
  ``Compactifications of F-Theory on Calabi--Yau Threefolds -- I,''
  Nucl.\ Phys.\  B {\bf 473}, 74 (1996)
  {\tt arXiv:hep-th/9602114}.



\bibitem{Bershadsky-all}
  M.~Bershadsky, K.~A.~Intriligator, S.~Kachru, D.~R.~Morrison, V.~Sadov and C.~Vafa,
  ``Geometric singularities and enhanced gauge symmetries,''
  Nucl.\ Phys.\  B {\bf 481}, 215 (1996)
  {\tt arXiv:hep-th/9605200}.

\bibitem{Katz-Vafa}
  S.~H.~Katz and C.~Vafa,
  ``Matter from geometry,''
  Nucl.\ Phys.\  B {\bf 497}, 146 (1997)
  {\tt arXiv:hep-th/9606086}.



\bibitem {Grassi-Morrison}
A.~Grassi, D.~R.~Morrison, ``Group representations and the Euler characteristic of elliptically fibered Calabi-Yau threefolds'',
 J.  Algebraic Geom.  12 (2003), 321-356
{\tt 	arXiv:math/0005196}.

\bibitem {Grassi-Morrison-2}
A.~Grassi, D.~R.~Morrison, ``Anomalies and the Euler characteristic of elliptically fibered Calabi-Yau threefolds,''
{\it to appear}.

\bibitem{Erler}
  J.~Erler,
  ``Anomaly Cancellation In Six-Dimensions,''
  J.\ Math.\ Phys.\  {\bf 35}, 1819 (1994)
  {\tt arXiv:hep-th/9304104}.

\bibitem{perrin}
Daniel Perrin, ``Algebraic geometry: an introduction,'' Springer, 2008.

\bibitem{Clay-Vafa}
  S.~Cecotti, C.~Cordova, J.~J.~Heckman and C.~Vafa,
  ``T-Branes and Monodromy,''
  {\tt arXiv:1010.5780 [hep-th]}.



\bibitem{Dienes-mr}
  K.~R.~Dienes, J.~March-Russell,
  ``Realizing higher level gauge symmetries in string theory: New embeddings for string GUTs,''
  Nucl.\ Phys.\  {\bf B479}, 113-172 (1996).
  {\tt hep-th/9604112}.

\bibitem{Dienes}
  K.~R.~Dienes,
  ``New constraints on SO(10) model building from string theory,''
  Nucl.\ Phys.\  {\bf B488}, 141-158 (1997).
  {\tt hep-ph/9606467}.

\bibitem{BV-global}
  M.~Bershadsky and C.~Vafa,
  ``Global anomalies and geometric engineering of critical theories in six
  dimensions,''
  arXiv:hep-th/9703167.

  
\bibitem{Elitzur}
  S.~Elitzur and V.~P.~Nair,
  ``Nonperturbative Anomalies In Higher Dimensions,''
  Nucl.\ Phys.\  B {\bf 243}, 205 (1984).

\bibitem{Cahn:1985wk}
  R.~N.~Cahn,
  ``Semisimple Lie Algebras And Their Representations,''
{\it  Menlo Park, Usa: Benjamin/cummings ( 1984) 158 P. ( Frontiers In Physics, 59)}.
Available online at 
{\tt http://phyweb.lbl.gov/\textasciitilde rncahn/www/liealgebras/book.html}


\end{thebibliography}
\end{document}